\documentclass[aps,prb,reprint,superscriptaddress]{revtex4-2}
\usepackage[utf8]{inputenc}
\usepackage{graphicx}
\usepackage{subfigure}
\usepackage{xcolor}

\newcommand{\vev}[1]{{\left< {#1} \right>}}

\newcommand{\prt}[1]{{\left( {#1} \right)}}

\newcommand{\be}{\begin{eqnarray}}
\newcommand{\ee}{\end{eqnarray}}

\newcommand{\wbe}{\begin{widetext}}
\newcommand{\wee}{\end{widetext}}

\usepackage{bbm}
\usepackage{amsmath,leftidx}
\usepackage{graphicx}
\usepackage{times}
\usepackage{CJK}
\usepackage{color,colortbl}
\usepackage[normalem]{ulem}

\begin{document}

\title{Berezinskii–Kosterlitz–Thouless transition from Neural Network Flows}

\author{Kwai-Kong Ng}
\email{kkng@thu.edu.tw}
\affiliation{Department of Applied Physics, \\
Tunghai University, Taichung 40704, Taiwan}

\author{Ching-Yu Huang}
\email{Corresponding author: cyhuangphy@thu.edu.tw}
\affiliation{Department of Applied Physics, \\
Tunghai University, Taichung 40704, Taiwan}

\author{Feng-Li Lin}
\email{Corresponding author: fengli.lin@gmail.com}
\affiliation{Department of Physics, \\
National Taiwan Normal University, Taipei, 11677, Taiwan}

\begin{abstract}
We adopt the neural network flow (NN flow) method to study the Berezinskii–Kosterlitz–Thouless (BKT) phase transitions of the 2-dimensional $q$-state clock model with $q\ge 4$. The NN flow consists of a sequence of the same units to proceed the flow. This unit is a variational autoencoder (VAE) trained by the data of Monte-Carlo configurations in the way of unsupervised learning. To gauge the difference among the ensembles of Monte-Carlo configurations at different temperatures and the uniqueness of the ensemble of NN-flowed states, we adopt the Jesen-Shannon divergence (JSD) as the information-distance measure ``thermometer". This JSD thermometer compares the probability distribution functions of the mean magnetization of two ensembles of states. Our results show that the NN flow will flow an arbitrary spin state to some state in a fixed-point ensemble of states. The corresponding JSD of the fixed-point ensemble takes a unique profile with peculiar features, which can help to identify the critical temperatures of BKT phase transitions of the underlying Monte-Carlo configurations. 
\end{abstract}

\date{\today}

\maketitle


\section{Introduction}

Machine learning is a powerful tool in extracting relevant features from big data and then classifying the data according to these learned features, and has a wide range of applications to physical sciences  \cite{troyer1,Hush:2017,Cai:2018,Torlai:2018,exotic1,fermionsign1,Carleo_2019,Liao:2021vec,Kuo:2021qtt}. When applying to physics study, the thermal states or quantum states of many-body systems are the natural arena to exploit the power of machine learning in dealing with big data. Indeed, there have been many studies of identifying the phase transitions and critical states by machine learning, e.g.,  \cite{melkonat1,spinlic2,manybodyloc,looptopography,huber1,Nieuwenburg:2018,Alexandrou:2019hgt,Kming2021UnsupervisedML,Iso:2018yqu,giataganas1,Koch:2019fxy,Giataganas:2021jqm,Suchsland:2018,glassy1,Ohtsuki:2019qnk,ronhovde2011detecting,Nicoli_2020,review1}. Most studies are based on the so-called supervised learning by training the neural network with the labeled data, e.g., labeling the thermal states by temperatures and then using the trained machine to classify the unknown states and identify the phase transition point. There are two challenges to such studies.  One is to discover new physics by machine learning, such as the new critical states. With supervised learning, the answers, e.g., critical states, are already encoded in the training set. It is desirable for the machine to be able to identify the target states without specifying them in advance in the training stage.  This indeed is the practice of the unsupervised machine learning, by which the machine is trained with unlabeled data to find out the posteriors of the data set encoding the information of the relevant features. 

There are many approaches to uncover the phase structures through unsupervised learning, for example \cite{Nieuwenburg:2018,Alexandrou:2019hgt,Kming2021UnsupervisedML,Iso:2018yqu,giataganas1,Koch:2019fxy,Giataganas:2021jqm}. Among them, the approach of the so-called neural network flow (NN flow) \cite{Iso:2018yqu,giataganas1,Koch:2019fxy,Giataganas:2021jqm} is more intriguing. Instead of training the machine to classify the thermal states, one trains a unit of flow with the unlabeled data and then flows the input states subsequently by the same trained unit. Surprisingly, the flowed states will approach a fixed-point ensemble of states, by which one can identify the critical phase transition by some thermometer. The validity of the NN flow method has been demonstrated with its application to $q=2,3,4$ Potts and clock spin models with various frameworks of unsupervised machine structure, such as restricted Boltzmann machine (RBM) and (variational) autoencoder (AE/VAE) \cite{Giataganas:2021jqm}. 
The success of NN flow reminds its similarity to the renormalization group flow (RG flow) which flows any state to the critical state by sequential scaling/rescaling procedures, and the unit of NN flow plays a similar role of the scaling/rescaling operation. Despite that, the underlying mechanism for the success of NN flow is not completely clear.

Another challenge is to apply the machine learning technique to identify the topological phase transition, such as the Berezinskii-Kosterlitz-Thouless (BKT) transition \cite{Kosterlitz:1973xp,Berezinskii1,Berezinskii2,kosterlitz2}. Unlike the usual Landau-Ginzburg type phase transition, there is no local order parameter to characterize the BKT phase transition, thus in some sense the BKT phase transition is continuous, or the so-called ``topological". One may need to adopt the non-local observables such as the vortex condensation to distinguish the BKT phase from the non-BKT ones.  In terms of pattern recognition from the machine learning point of view, this means that the relevant features of the BKT phase transition should be less obvious than the Landau-Ginzburg ones. The BKT phase transition was first studied for the XY model. A family of its discrete versions, called the $q$-state clock model with $q \ge 5$ can also display the BKT phase \cite{Jose:1977gm,ORTIZ2012780}.  
The BKT phase appears in the XY model whenever the temperature $T$ is below the critical temperature. However, for the above clock models, the BKT phase appears when $T_1\le T \le T_2$. Below $T_1$, the model is in the ferromagnetic ($Z_q$ broken) phase, and above $T_2$ the paramagnetic (disordered) phase. Thus, there are the BKT phase transitions at both $T_1$ and $T_2$, which are not easy to identify precisely because of the lack of local order parameter of Landau-Ginzburg type even by numerical methods \cite{Tomita2002ProbabilityChangingCA,Krvcmar2016PhaseTO,Chatelain2014DMRGSO,Vanderstraeten:2019frg,Li2020CriticalPO}. Despite that, there exists the so-called {\it extended university} for the clock model of $q>4$ by which the continuous symmetry emerges for $T>T_{eu}$ with the $T_{eu}$ the onset temperature for the collapse of thermodynamic observables. The existence of $T_{eu}$ turns the phase transition at $T_2$ of the $q=5,6$ clock models into the non-BKT one \cite{Lapilli_2006}.

Despite the intriguing feature of topological phase transitions and the difficulty of characterization, there are some machine learning studies that identify BKT phases for clock models based on the supervised learning scheme \cite{Shiina2020MachineLearningSO,Tomita_2020}. In this work, we apply the NN flow method to identify the BKT phase transitions for the clock models. In our previous work \cite{Giataganas:2021jqm}, we have studied the NN flows of the clock models with $q\le 4$, of which the phase transitions are of Landau-Ginzburg type. Regarding the $q\ge 5$ cases, we have also done some trials during the working period for \cite{Giataganas:2021jqm}, but cannot identify the BKT phase transitions. There are two reasons for this failure. One is due to inaccurate training data for BKT states by our chosen Monte-Carlo method. The other is due to an inaccurate machine learning thermometer, which cannot give a precise temperature reading of the NN-flowed configurations in the low-temperature regime because of the less relevant features in such regime.  
In this work, we have improved both for the implementation of NN flow for BKT phase transitions. We simply adopt a more accurate Monte Carlo scheme so that the simulated states can exhibit the BKT phase transitions and serve as the training data for the unit of NN flow. 
For characterizing the configurations, we give up the machine-learning thermometer, which is hard to train to be accurate enough for the low temperature regime. Instead, we adopt the information-distance quantity, the so-called Jensen-Shannon divergence (JSD), which is a generalization of the Kullback-Leibler divergence, to measure the difference between two ensembles of thermal states at different temperatures. Therefore, we can use the Monte-Carlo states of a given temperature as the gauge ensemble to compare with the ensemble of the NN-flowed states. 

In this way, we first see that the NN-flowed ensembles obtained from the different temperature ensembles of Monte-Carlo simulated configurations all yield the same information-distance measure profile. This implies that all the NN-flowed states belong to the same ensemble of states, which we call the fixed-point ensemble (of states). Next, the JSD profile shows peculiar features at phase transitions to help identify critical temperatures $T_{1,2}$. For example,  the information-distance measure shows a sharp drop near $T_1$, and a local minimum near $T_2$.  Our results show that the NN flow can flow an arbitrary state to a state in the fixed-point ensemble, and the pattern of its JSD can be used to identify the critical temperatures of BKT phase transitions. Since the JSD shows a minimum near $T_2$, we may expect that the NN-flowed states capture the essence of the critical states at $T_2$. However, this is not the case, as we will discuss later.

The remainder of this paper is organized as follows. In section \ref{sec:2} we review the clock models, BKT phase transitions and describe our Monte-Carlo method and the phase diagrams of the simulated results. In section \ref{sec:3} we review the method of NN flow. In section \ref{sec:4} we introduce the information-distance ``thermometer" for measuring the temperatures of thermal ensemble states. In section \ref{sec:5}, we show the results of the NN flows of the $q=2,4,5,6,7,8$ clock models on $20\times 20$ and $40\times 40$ lattices, including the patterns of JSDs and their features for identifying the (non-)BKT phase transitions. We also examine whether the fixed-point ensemble states bear some physical properties of the BKT phase.  We conclude our paper in section \ref{sec:6}. In Appendix \ref{app_a} we show more on the BKT phase diagrams of our Monte-Carlo configurations. 


\begin{figure}
 \includegraphics[width=0.55\textwidth]{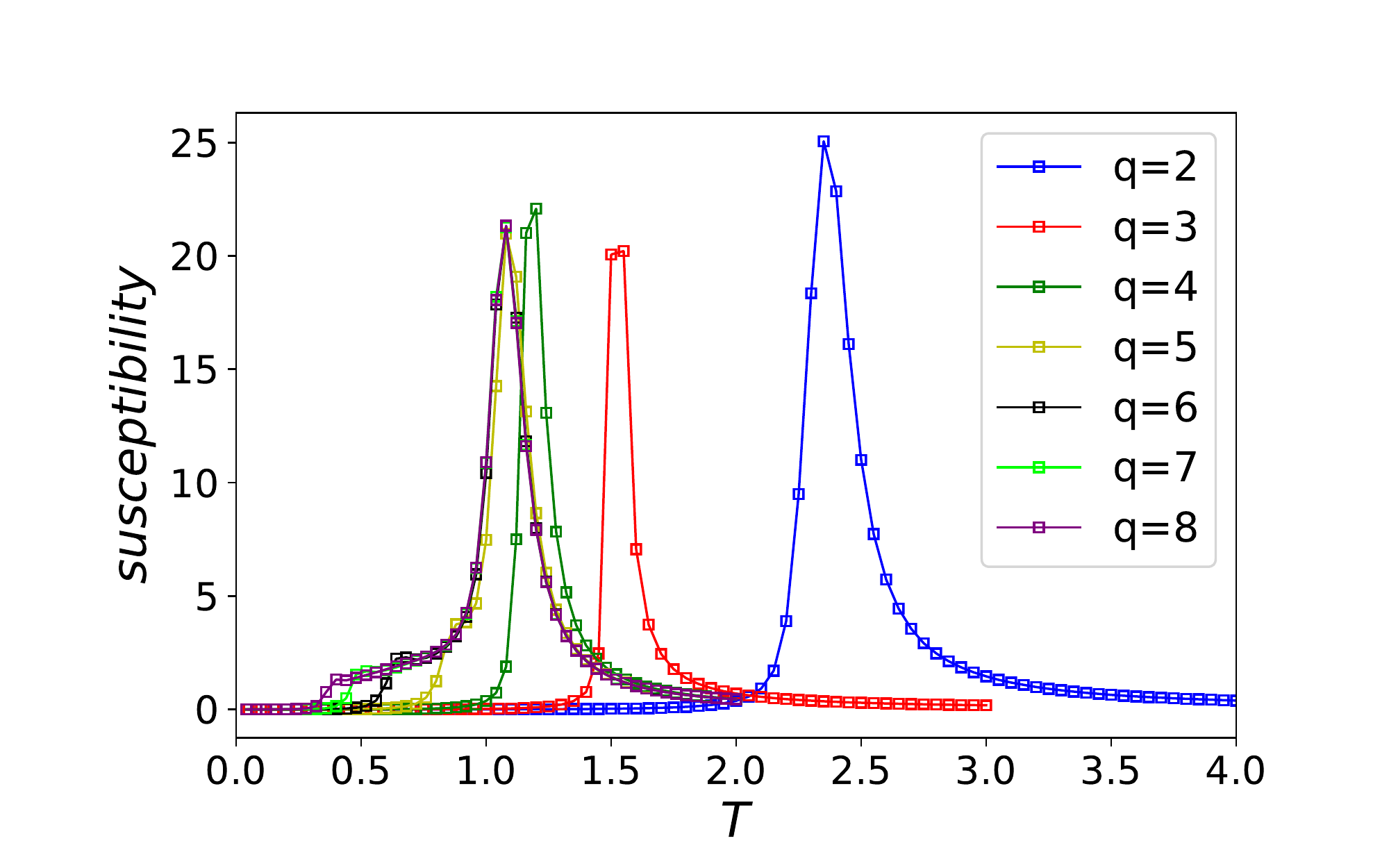}
 \caption{\small Phase diagrams for the $q$-state clock model based on the temperature profiles of magnetic susceptibility, which are obtained from Monte-Carlo simulations on a $40 \times 40$ lattice.
 }\label{fig:L40_chi}
\end{figure}

\section{Mote-Carol simulation of q-state clock model and Berezinskii–Kosterlitz–Thouless transition}\label{sec:2}

We study the $q$-state clock model on the square lattice,  which is described by the following Hamiltonian 
\be 
\label{hamilton}
{\cal H} =-J \sum_\vev{ij} \cos(\theta_i- \theta_j)~,
\ee
where the $q$-state spin on the site $i$ is characterized by the Potts variable $\theta_i=2\pi s_i/q$  with $s_i=0,1,\cdots,q-1$, and $\vev{ij}$'s denote the nearest neighbors that interact by the ferromagnetic coupling $J>0$. 
The thermodynamic behaviors of the model and the associated  phase transitions are encoded in the following partition function at finite temperature $T$,
\be
Z_N=\sum_{\textrm{all possible } \theta_i, \theta_j} \exp\prt{ {J \over k_B T} \sum_\vev{ij}  \cos(\theta_i- \theta_j) }\;.
\ee
In the following we set $J=1$ as the energy unit. The q-state clock model has been extensively investigated both analytically~\cite{ORTIZ2012780} and numerically~\cite{Tomita2002ProbabilityChangingCA,Lapilli2006UniversalityAF,Chatelain2014DMRGSO,Krvcmar2016PhaseTO,Vanderstraeten:2019frg,Li2020CriticalPO}.
The model is exactly solvable for $q=2,3,4$ \cite{pottsmodel}. For $q=2,3$ they reduce to the Ising and the $3$-state Potts models, respectively; and for $q=4$ it can be mapped into a double copy of the Ising model. Therefore, their phase transitions are second order of Landau-Ginzburg types. For $q\ge 5$, the model starts to display the BKT phases sandwiched by a low temperature ferromagnetic ($Z_q$) broken phase and a high temperature paramagnetic (disordered) phase. Thus, there are two continuous BKT phase transitions at temperature $T_1$ and $T_2\; (>T_1)$. 
In \cite{Miyajima_2021}, the machine learning technique and finite-size scaling are adopted to determine the critical temperatures of the $q=8$ clock model, which are $T_1= 0.410$ and $T_2=0.921$ in the thermodynamic limit.
As $q\rightarrow \infty$, i.e., a planar rotor, the model is equivalent to the XY model for which the BKT phase extends throughout the low temperature regime, that is, $T_1=0$.  The BKT phase transitions are notoriously difficult to identify \cite{Lapilli2006UniversalityAF,Tomita2002ProbabilityChangingCA,Krvcmar2016PhaseTO,Chatelain2014DMRGSO,Vanderstraeten:2019frg,Li2020CriticalPO}, as there is no Landau-Ginzburg-type order parameter to characterize them. This poses the challenge to identify the BKT phase transitions by some novel way, and then motivate this work to resolve it by the NN flow.  


To proceed the NN flow for identifying the BKT phase transitions of the clock models, we need to first produce the spin configurations as the training data to the neural network, i.e., the variational autoencoder (VAE) for the current work. We will implement the Monte-Carlo method to simulate the spin configurations of the training set for various temperatures.

In the Monte Carlo simulation, it is known that, near the critical temperature, single-flip updating approach will encounter the problem of critical slowing down, in which the autocorrelation time, i.e. the time for reaching independent configurations, becomes very large. To tackle this problem, Swendsen and Wang \cite{PhysRevLett.58.86} proposed a cluster updating scheme that connected sites of same spins are grouped into a cluster according to some assigned probabilities. In this way spins are flipped altogether inside the same cluster. Later, Wolff \cite{PhysRevLett.62.361} modified the cluster algorithm and extended it to multi-component spins such that the spins are first projected into a randomly chosen direction before the Swednsen and Wang algorithm is carried out. Adopting the Wolff algorithm, we are able to obtain accurate Monte Carlo configurations for the NN flow. We generate 1000 configurations for each temperature with 50 to 80 different temperatures for each $q$ value. In order to investigate the size effect, simulations of both lattice sizes $L=20$ and $L=40$ are also carried out.

In Fig.~\ref{fig:L40_chi} we show the magnetic susceptibility of the clock models for $q=2,3,4,5,6,7,8$, which are obtained from our Monte-Carlo simulated configurations. The phase diagrams based on the temperature profiles of magnetization, energy, and heat capacity can be found in Fig. \ref{fig:MC_data} of the Appendix \ref{app_a}. For simplicity, we will focus only on Fig. ~\ref{fig:L40_chi} for discussion. From Fig. ~\ref{fig:L40_chi} we can read the critical temperatures. We can see that the phase diagrams (or temperature profiles of magnetic susceptibility) move towards the low-temperature regime as $q$ increases. 
This is in part why our previous tests of the NN flow for $q\ge 5$ fail because the resolution of our machine-trained thermometer is poor in the low-temperature regime. We can also see that for $q\ge 5$ models, there is a sharp peak near $T_2$ but only a minor jump near $T_1$. Despite that, we will see that our information-distance thermometer will give clearer features near $T_{1,2}$. 

Also, in Fig. \ref{fig:L40_chi} the lattice size is $40\times 40$. Due to the topological nature of the BKT phase transitions, the critical temperatures $T_1$ and $T_2$ depends more sensitively on the lattice size than the Landau-Ginzburg ones, for example see \cite{Tomita2002ProbabilityChangingCA} for the study of finite-size effect of $q=6$ clock model.  Our goal here is not to find the true (theoretical) $T_{1,2}$ in the large size limit, but to check whether the NN flow can be adopted to identify the $T_{1,2}$ of the Monte-Carlo simulations. It will be much more challenging for machine learning to identify the theoretical critical temperatures using the Monte-Carlo training set on the finite-size lattice. For simplicity, we will not specify the theoretical values of $T_{1,2}$. Later on, we will just examine the information-distance measure of the NN-flowed states to the Monte-Carlo ones. In this way, we can examine the validity of the NN-flow method.

\begin{figure}
 \includegraphics[width=0.35\textwidth]{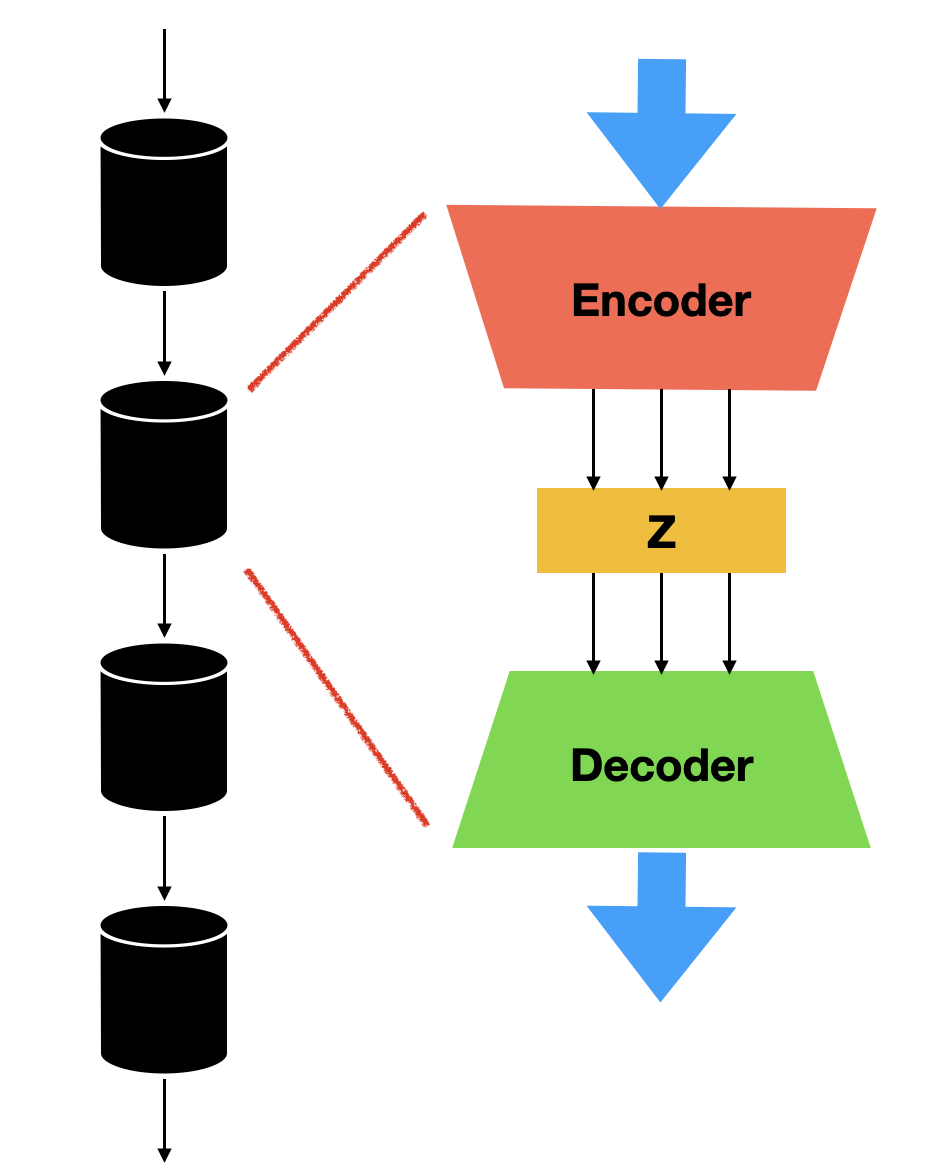}
 \caption{\small Schematic structure of the neural network flow (Left) with the substructure of each unit, i.e., an (variational) autoencoder shown in the right. In this work, we consider VAE so that the hidden layer vector $z$ is stochastic with Gaussian distributions.}
 \label{fig:NN_flow}
\end{figure}

\begin{figure}
 \includegraphics[width=0.35\textwidth]{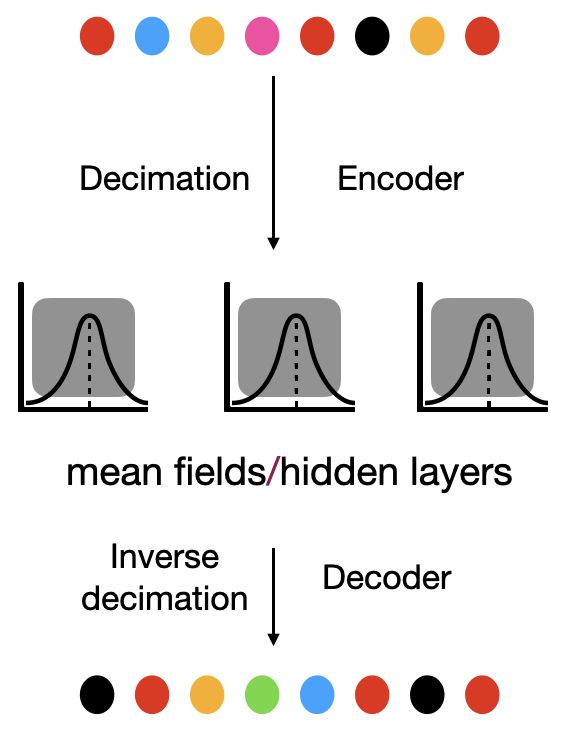}
 \caption{\small  Naive analogue between RG flow and neural network flow. The decimation of the renormalization flow is in analogy to the encoder and its inverse to the decoder. Therefore, the stochastic hidden layer plays the role of mean field in the renormalization flow. As a result, the different choices of the stochastic distribution for the hidden layer may correspond to the different mean field and yield different fixed-point states. }
 \label{fig:RG_vs_NN}
\end{figure}

\section{Neural Network Flow}\label{sec:3}
\label{sec:NNflow}

In this section, we review our setup for the NN flow, which is originally proposed in \cite{Iso:2018yqu} (see also \cite{giataganas1}), and then generalized in our previous work \cite{Giataganas:2021jqm} by adopting not only the restricted Boltzmann machine (RBM) but also the autoencoder (AE) and the variational autoencoder (VAE) as the unit of NN flow. In this work, we will simply adopt VAE since it yields more efficient and stable results from our own experiences. 
 
The NN flow scheme is shown in the left part of Fig. \ref{fig:NN_flow}, which consists of a sequential flow through the NN (i.e., VAE adopted in this work) units. The detail of the VAE unit is shown in the right part of Fig. \ref{fig:NN_flow}. The procedure of the NN flow goes as follows. First, we train the VAE with Monte-Carlo simulated configurations of the clock models. For each temperature $T$, we prepare about $2000$ configurations as the training set, and the size of the temperature bin is $0.1$ with $0\le T \le 4$. The structure of VAE is very simple. It consists of an encoder and a decoder, as indicated in the right part of Fig. \ref{fig:NN_flow}. The encoder  consists of an input layer to which the input spin configuration is preprocessed by proper normalization as the common practice at the training stage of typical machine learning, and a hidden/latent layer whose elements are random variables of unit Gaussian. One can also add an intermediate layer before the hidden layer. The sizes of the intermediate and hidden layers can be adjusted accordingly to optimize the training. The decoder then takes the hidden layer as the input, either or not goes through an intermediate layer, and then gives the output layer with the same size as the input layer of the encoder. The mean square error between the input layer and the output layer can be calculated as the penalty for optimizing the machine structure. Usually, we do not need to have very good training of the VAE to give some flexibility to flow the state. Otherwise, if the VAE is well-trained, then the output state will be perfectly the input state, and there will be no flow. The detailed machine structure of VAE and the training accuracy are almost the same as the one used in \cite{Giataganas:2021jqm}, so that we will not mention here but refer to \cite{Giataganas:2021jqm} for the details.

In some sense, VAE has the structure similar to a single step of renormalization group (RG) flow, as depicted in Fig. \ref{fig:RG_vs_NN}. The encoder compresses the input data by some truncation, similar to the coarse-graining by decimation. This procedure is used to try to approach a mean-field state, which in NN flow is the Gaussian hidden layer. The encoder then proceeds the hidden layer to the original size of the input layer, which is quite similar to performing the inverse decimation in the RG flow. It is interesting to examine how good this analogue can be in the current study of the BKT phase transitions. 

After training the VAE, we then use it as the unit of NN flow. The NN flow then passes an arbitrary Monte-Carlo configuration into a sequence of unit VAEs. All the VAEs have the same bias and weights determined at the training stage. The procedure of NN flow is similar to a sequence of decimation/inverse decimation steps in the RG flow to flow a UV state to the IR fixed-point state. Typically, after a few steps, the NN flow will also yield a fixed-point state. However, unlike the RG flow, we will see that the ensemble of fixed-point states bears no characteristic of criticality, such as power-law behaviors of correlation functions. Despite that, our information-distance thermometer can still indicate peculiar features near $T_{1,2}$.

We now motivate the need to construct a ``thermometer"  based on  information-distance measure to gauge the NN-flowed states, and to use it to identify the BKT phase transitions. For the Landau-Ginzburg type phase transition, we can indeed calculate the order parameter such as the magnetization of the Ising model (or equivalently the $2$-state clock model \cite{pottsmodel}) and compare with the one by Monte-Carlo method. Or we can measure the temperature of the NN-flowed states by using a machine-learning thermometer. 
We can train the thermometer machine with the Monte-Carlo simulated data set. In fact, this is what has been done in our previous work \cite{Giataganas:2021jqm}. However, this kind of thermometer is not very accurate at low temperature regime. This is expected as the low temperature phase is the ordered phase with less intrinsic features, thus more difficult to train. When studying BKT phase, the above techniques become less useful, because there is no local order parameter such as magnetization for BKT phase transition, and especially the BKT critical temperatures tend to locate at the low temperature regime. This is why we need to adopt different techniques to characterize the NN-flowed states to extract their BKT features. This will be discussed in the next section.

\begin{figure*}[t!]
\includegraphics[angle=0,width=1.0 \textwidth]{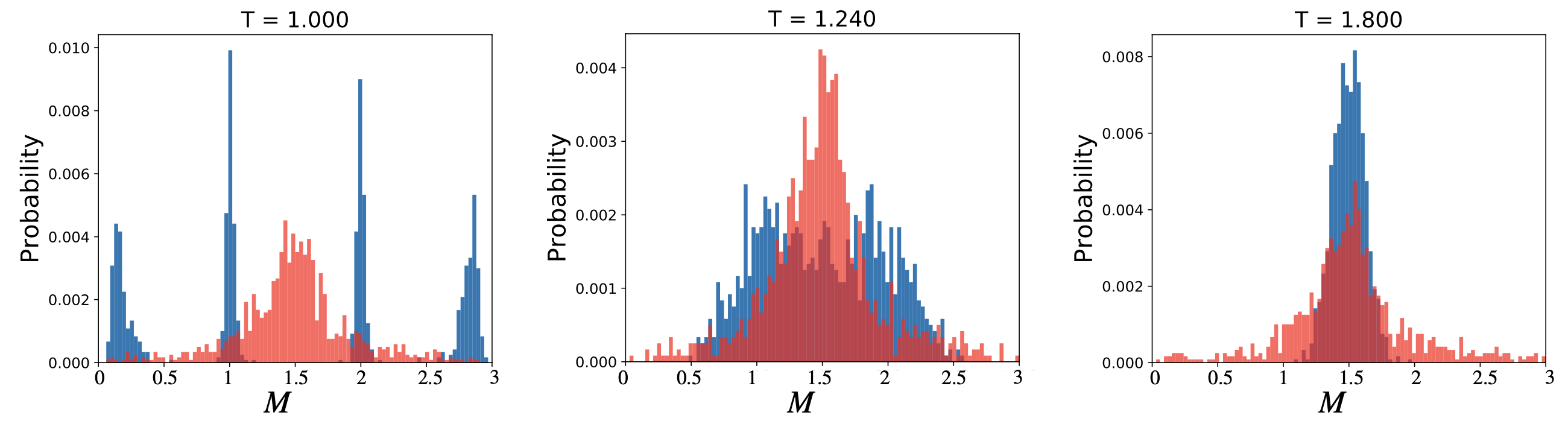}
\caption{\small The probability density functions (PDF) for the mean magnetization $M_n$. Blue/red color: the PDFs for the temperature ensembles of Monte-Carlo simulated configurations of $q=4$ clock model on $20 \times 20$ lattice at $T=1$ (below $T_c$), $T=1.24$ (near $T_c$) and $T=1.8$ (above $T_c$). Orange/red color: the PDFs of the corresponding NN-flowed ensemble of the top-row ones after 100 iterations. It shows that the NN-flowed states approach some states in a  fixed-point ensemble of states. Note that the red-color part is the overlap of the blue and orange ones.
\label{fig:M_dis_q4}}
\end{figure*}

\begin{figure}[] 
\centering
    \subfigure[  ]{
        \includegraphics[width=0.43\textwidth]{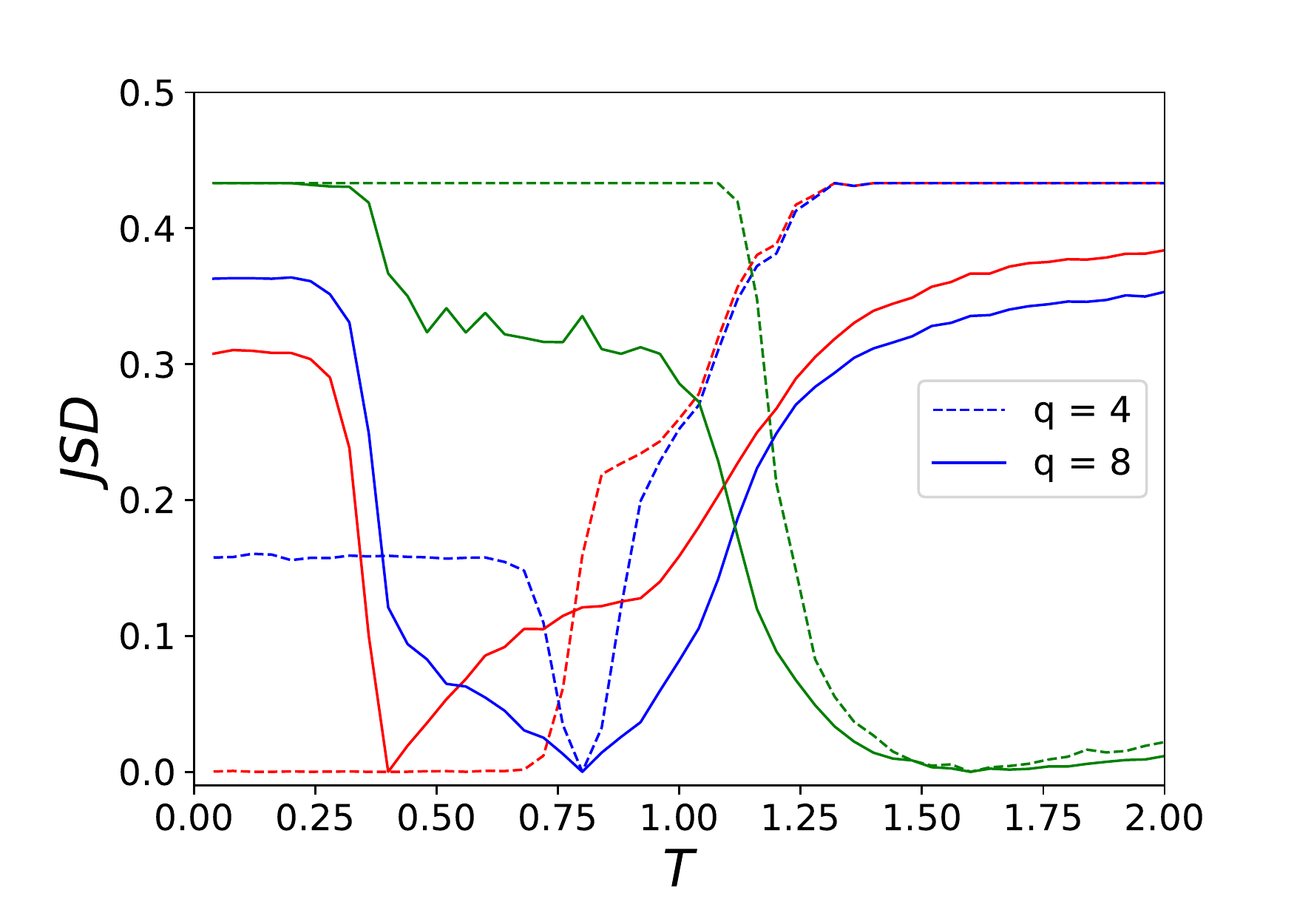} }
    \subfigure[ ]{
        \includegraphics[width=0.43\textwidth]{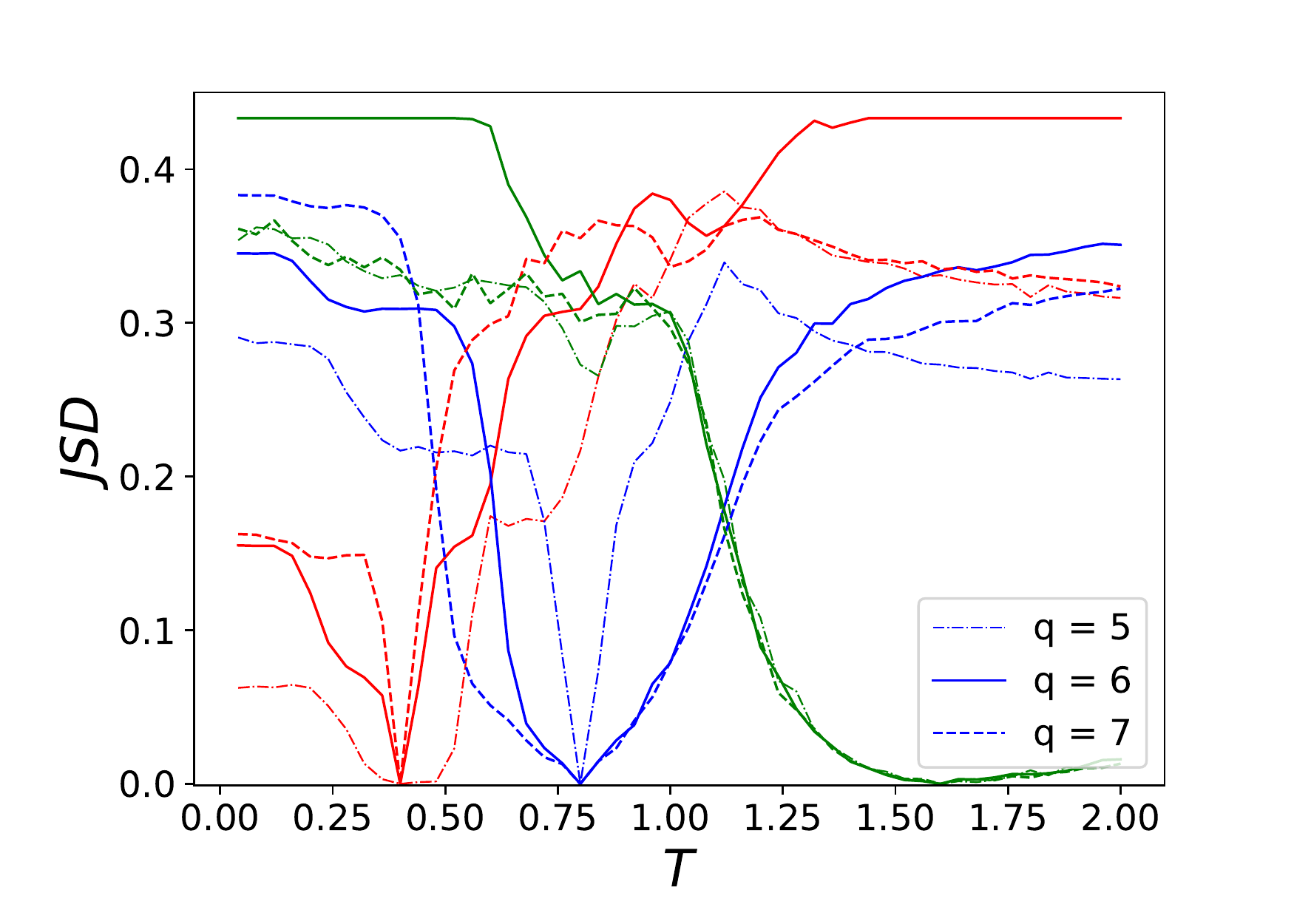} }
\caption{\small Jensen–Shannon divergence ${\rm JSD}[P_{T_p}|| Q_T]$ of the (a) $q=4,8$  and   (b) $q=5,6,7$  (see the labels) clock models on a $40\times 40$ lattice. Here $T$ runs all the temperature bins in $0<T<4$, and $T_p$ runs for three specific temperatures: $T_{\rm low}=0.4$ (red lines), $T_{\rm mid}=0.8$ (blue lines) and $T_{\rm high}=1.6$ (green lines). The JSDs can serve as a good ``thermometer" because the minimal of the JSDs sit right on the corresponding $T_p$.}
\label{fig:JSD_demo}
\end{figure}

\section{Information-distance measure as thermometer} 
\label{sec:4}

Instead of training an additional machine-learning type thermometer, we can in fact use the temperature ensembles of the Monte-Carlo simulated configurations to gauge NN-flowed states and its associated ensemble. In this sense, we would like to measure the difference between two ensembles of thermal states. Thus, one can invoke some information-distance measures such as the root-mean-square distance, the Kullback-Leibler divergence (KLD), or the Jensen-Shannon divergence (JSD). Given two (discrete) probability density functions (PDFs) denoted by $P$ and $Q$, the KLD is given by
\be
{\rm KL}[P||Q] =-\sum_n P(n) \ln {Q(n)\over P(n)}\;.
\ee 
KLD is a very common information-distance measure, also known as relative entropy.  However, it is not well-defined if $Q(n)$ vanishes for some $n$. Therefore, a more well-defined measure is JSD, which is given by
\be 
\label{JSD}
{\rm JSD}[P || Q]={\frac  {1}{2}} {\rm KL}[P || R]+{\frac  {1}{2}} {\rm KL}[Q || R]
\ee
where $R = \frac{1}{2}(P+Q)$. It is easy to see that JSD is well-defined even some of $P(n)$ or $Q(n)$ vanishes. Note that ${\rm KL}[P|| Q]$ and ${\rm JSD}[P|| Q]$ are positive definite and can measure the dissimilarity between distributions $P$ and $Q$. Thus, ${\rm JSD}[P || Q]$ becomes smaller if $P$ and $Q$ are more similar to each other, and vanishes if $P=Q$.

Recall that temperature can also be defined statistically. Thus, our information-distance measure in this sense plays the role of the ``thermometer". 
With JSD as the chosen information-distance measure, the next question is what kind of PDFs we will choose so that the resultant JSD can serve as a good ``thermometer". 
Since we would like to construct the PDFs for the ensemble of the thermal states, it is better to choose some specific property of the spin configurations and consider its PDF for a given ensemble. A natural property for the spin configurations is the mean magnetization, 
\be
M = {1\over N} \sum_{i=1}^{N} s_i 
\ee
where $N$ is the total number of the lattice sites, and $s_i=0,1, \cdots, q-1$ is the spin value on the site $i$. Thus, $M$ will be the random variable for an ensemble of a given temperature or its corresponding ensemble of the NN-flowed states. We will further discretize $M$ by proper rounding into bins denoted by $M_n$ with $n$ the integer index, so that its associated PDF will be just denoted by $P(n)$ or $Q(n)$ as the ones used for JSD. 

In Fig. \ref{fig:M_dis_q4} we show the PDFs for $M_n$ for the Monte Carlo configuration ensembles of the $q=4$ clock model at three different temperatures (in blue/red): $T=1$ (below $T_c$), $T=1.24$ (near $T_c$) and $T=1.8$ (above $T_c$), and also the PDFs of the corresponding ensemble states obtained by the NN flow (in orange/red color). We can see that the NN-flowed PDFs resemble each other. 
As we shall see later, this implies that the NN-flowed states approach some states in a fixed-point ensemble of states.

To demonstrate the ability of the JSD ``thermometer", in Fig. \ref{fig:JSD_demo} we plot the ${\rm JSD}[P_{T_p} || Q_T]$ for the $q=4,5,6,7,8$ clock models on a $40\times 40$ lattice. Here $Q_T$ is the PDF of the mean magnetization for the Monte-Carlo ensemble at temperature $T$ with $T$ running over all temperature bins with $0<T<4$, and $P_{T_p}$ is the one for some specific temperature $T_p$. 
In Fig. \ref{fig:JSD_demo} we choose three specific temperatures: $T_{\rm low}=0.4$, $T_{\rm mid}=0.8$ and $T_{\rm high}=1.6$ with the middle temperature $T_{\rm mid}$  in the BKT phase for the cases of $q>4$. 
We see that in most cases the JSD has its minimum sit right on $T=T_p$. This indicates that the JSD can indeed serve as a good ``thermometer" for the clock models considered in this work. 
The only exceptions are the JSDs in the low temperature regime of the $q\le 4$ cases, e.g., the one of $q=4$ at $T_{low}$ in Fig. \ref{fig:JSD_demo}, for which we find that the corresponding JSDs are almost zero in an extended regime at low temperature. It indicates that the thermal fluctuations in these low temperature ordered phases are too small to be resolved by JSD.

\section{The results of NN flow for BKT phase transitions}
\label{sec:5}

In this section, we will present the results of the NN flows for the $q=2,4,5,6,7,8$ clock models and discuss the properties of the NN-flowed states.

\subsection{$q=2$}

\begin{figure}[t!]
\includegraphics[angle=0,width=0.48 \textwidth]{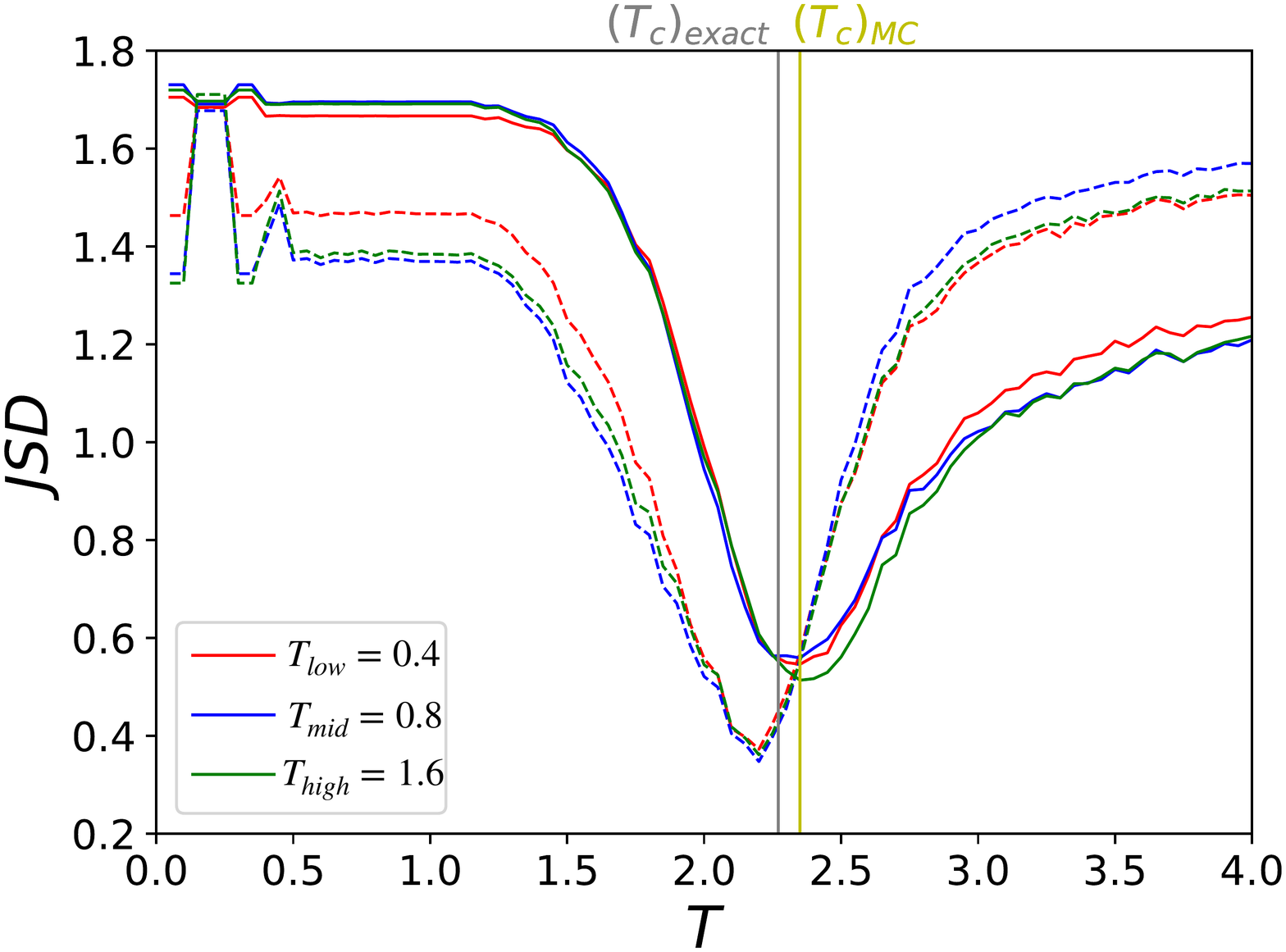}
\caption{ \small  NN-flowed ${\rm JSD}[P_{T_{\rm flow}}|| Q_T]$ of $q=2$ clock model on a $20\times 20$ lattice. Here $P_{T_{\rm flow}}$ is the PDF of the mean magnetization of NN-flowed states after $1$ iteration (dashes line) and $100$ iterations  (solid line) of the initial Monte-Carlo ensemble at temperature $T_{\rm flow}$, and $Q_T$ is the reference PDF based on Monte-Carlo ensemble at temperature $T$. For convenience, we choose $T_{\rm flow}=T_{\rm low}$ (red), $T_{\rm mid}$ (blue), and $T_{\rm high}$ (green) of Fig. \ref{fig:JSD_demo}. 
We see that there is a unique pattern for NN-flowed JSDs with its minimum sitting inside the narrow window of $[(T_c)_{exact},(T_c)_{MC}]$ with $(T_c)_{exact}=\frac{2}{\log{(1+\sqrt 2)}}=2.27$ (vertical gray line) and $(T_c)_{MC}=2.35$ (vertical light-yellow line) the theoretical and Monte-Carlo critical temperatures, respectively. }
\label{fig:q2_JSD_L20}
\end{figure}

\begin{figure}[t!] 
 \centering
    \subfigure[ ]{
        \includegraphics[width=0.47\textwidth]{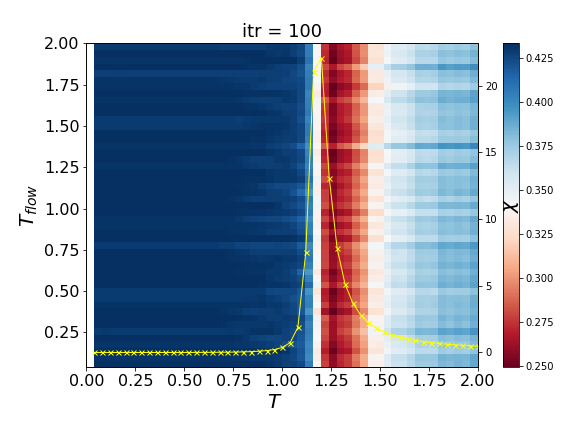} }
    \subfigure[ ]{
        \includegraphics[width=0.47\textwidth]{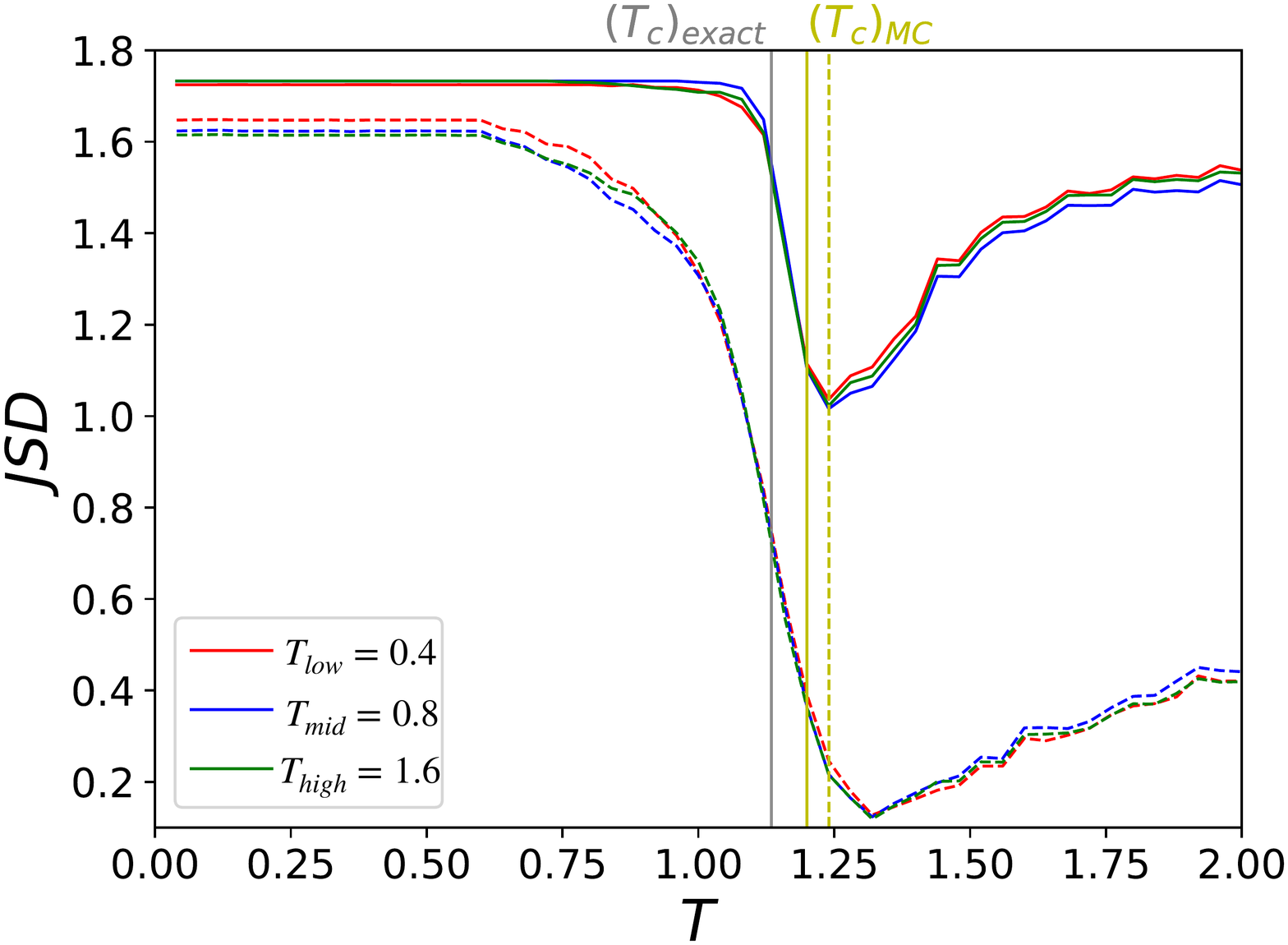} }
\caption{ 
(a) Color chart of ${\rm JSD}[P_{T_{\rm flow}}|| Q_T]$ of $T$-$T_{\rm flow}$ plane for the $q=4$ clock model on a $40\times 40$ lattice for the ensemble of NN-flowed states after $100$ iteration steps. The corresponding diagram of magnetic susceptibility (yellow line with decorated crosses) is also superimposed, which shows a peak near $T=(T_c)_{exact}$. 
(b) NN-flowed ${\rm JSD}[P_{T_{\rm flow}}|| Q_T]$ for $T_{\rm flow}=T_{\rm low}$ (red), $T_{\rm mid}$ (blue) and $T_{\rm high}$ (green) of Fig. \ref{fig:JSD_demo} on a $20\times 20$ lattice (dashed line) and  $40\times 40$ lattice (solid line).
We label $(T_c)_{exact}=\frac{1}{\log{(1+\sqrt 2)}}=1.13$ by the vertical gray line, and $(T_c)_{MC}=1.24\; (20\times 20 \; {\rm lattice})$, $1.20 \; (40\times 40 \; {\rm lattice})$ with the vertical solid light-yellow lines.  We see that the finite-size effect is relevant to improve the capability of using the minima of NN-flowed JSDs to identify the critical temperature. Our results again show that NN flows can reach a fixed-point JSD pattern with its minimum located at the critical temperature.}
\label{fig:q4_JSD}
\end{figure}

\begin{figure*}[t!] 
\centering
\subfigure[q=5]{
        \includegraphics[width=0.32\textwidth]{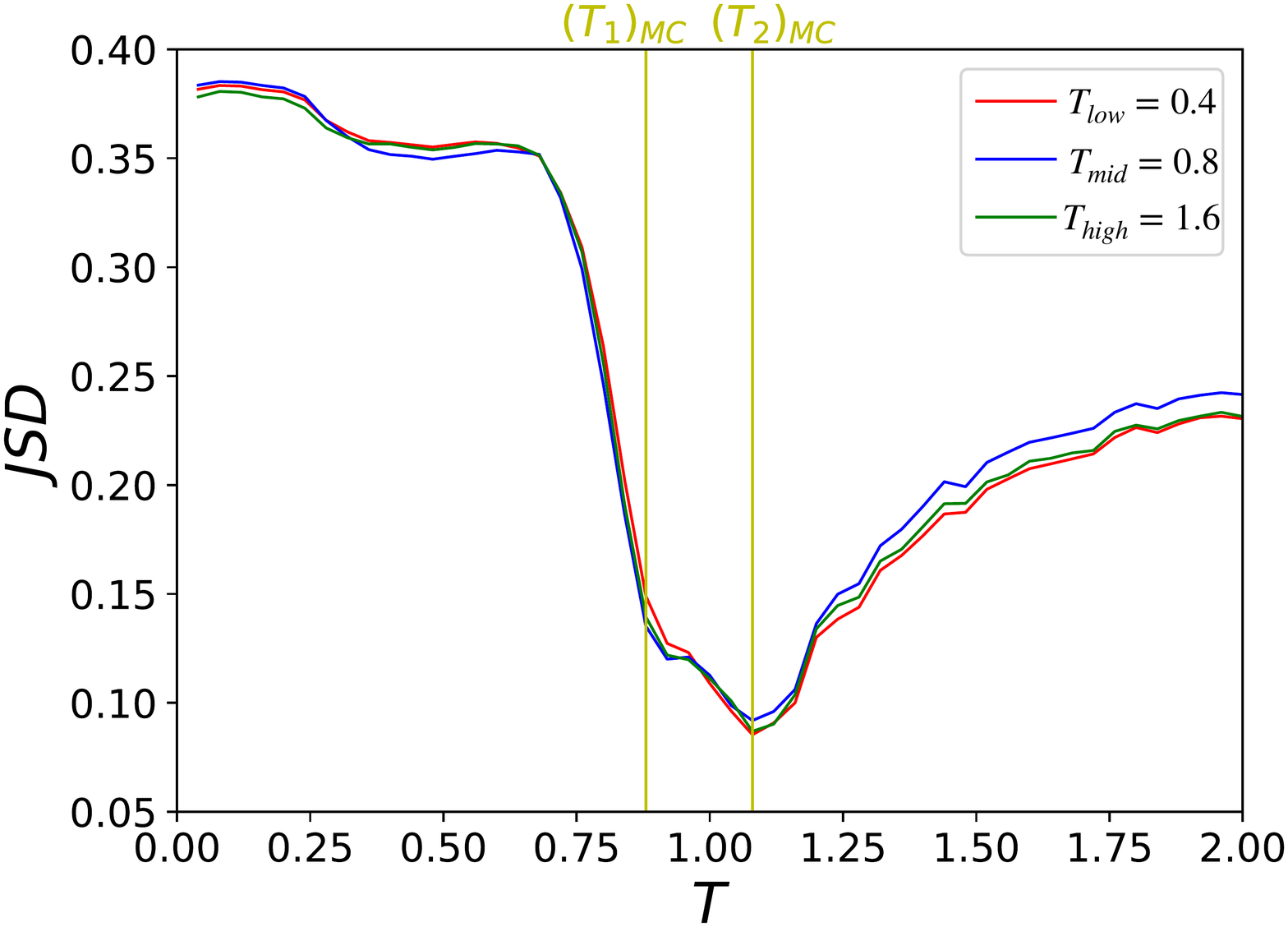}}
\subfigure[q=6]{
        \includegraphics[width=0.32\textwidth]{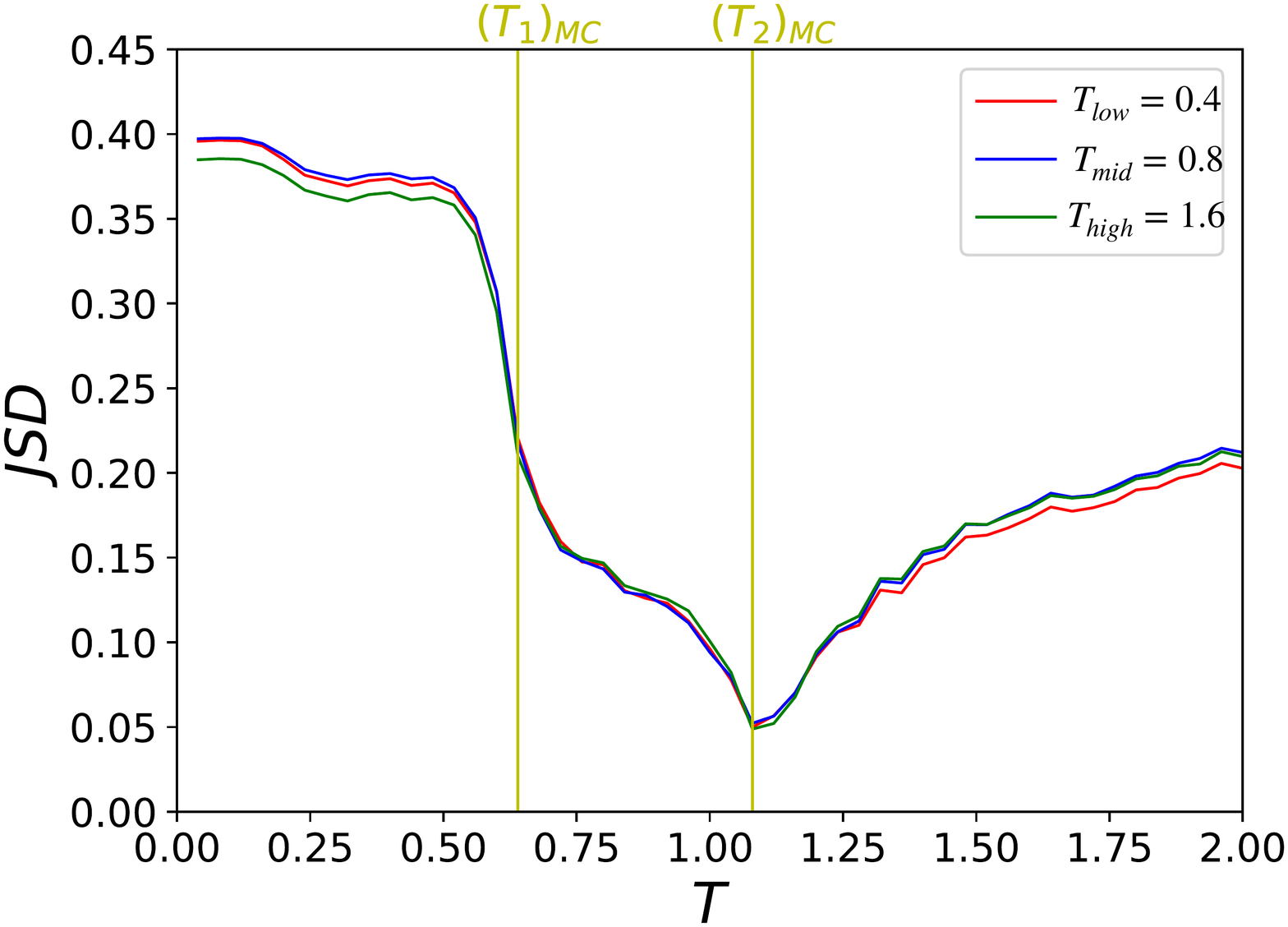}}
\subfigure[q=7]{
        \includegraphics[width=0.32\textwidth]{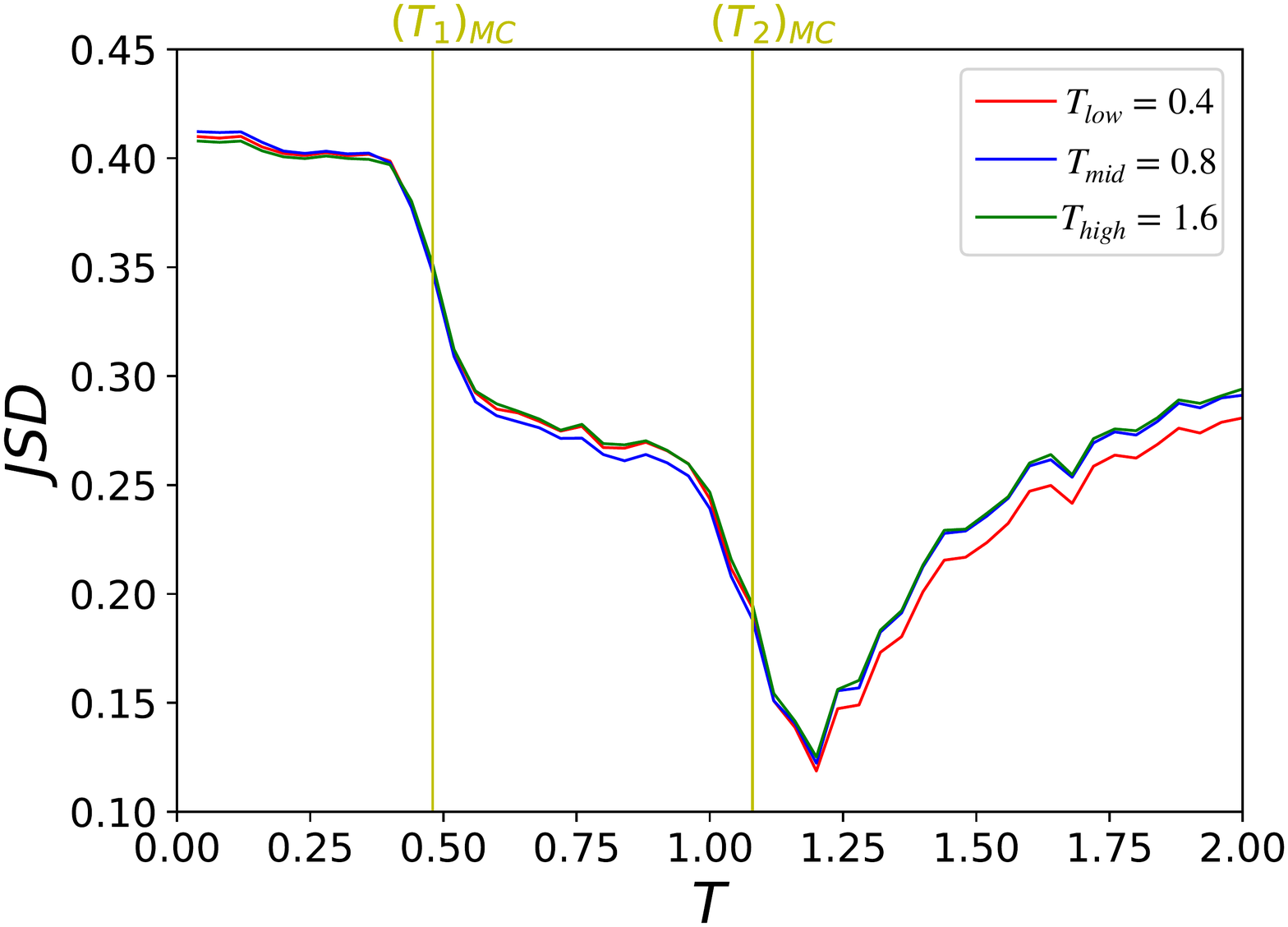}}
\caption{ 
\small NN-flowed ${\rm JSD}[P_{T_{\rm flow}}|| Q_T]$ after 100 iterations for $T_{\rm flow}=T_{\rm low}$ (red), $T_{\rm mid}$ (blue) and $T_{\rm high}$ (green) of Fig. \ref{fig:JSD_demo} on a $40\times 40$ lattice for the clock models of $q=$ (a) 5, (b) 6 and (c) 7. The critical temperatures $(T_{1,2})_{MC}$ of Monte-Carlo simulation are indicated by the vertical light-yellow lines in each subfigure. 
A peculiar plateau of the universal JSDs is roughly overlapped with the regime of the BKT phase interval $[(T_1)_{MC}, (T_2)_{MC}]$. Moreover, the JSD minima sit right at $T=(T_2)_{MC}$ in (a) and (b), but a bit higher than $T=(T_2)_{MC}$ in (c).   }
\label{fig:q5_q6_q7flow}
\end{figure*}

We first consider the $q=2$ clock model on the $20\times 20$ lattice as a benchmarking test to check if the JSD of mean magnetization can be used as a kind of ``thermometer" to gauge the ensembles of states. As mentioned, the $q=2$ clock model is equivalent to the Ising model with a second-order phase transition from the low temperature ordered phase to the high-temperature disordered one.  
We will consider the NN-flowed ${\rm JSD}[P_{T_{\rm flow}}|| Q_T]$ where $P_{T_{\rm flow}}$ is the PDF of the mean magnetization of NN-flowed states after $100$ iterations of the initial Monte-Carlo ensemble at temperature $T_{\rm flow}$, and $Q_T$ is again the Monte-Carlo one at temperature $T$ with $T$ running over $0<T<4$. For convenience, we choose $T_{\rm flow}=T_{\rm low}$, $T_{\rm mid}$, and $T_{\rm high}$ of Fig. \ref{fig:JSD_demo}. 
We expect that the NN-flowed JSDs should show a unique pattern associated with a fixed-point ensemble of states at critical temperature, thus to be consistent with the results obtained in \cite{Giataganas:2021jqm}. 
This is indeed the case as shown in Fig. \ref{fig:q2_JSD_L20}, as we see that the minimum of the universal NN-flowed JSD sits inside the narrow window of $[(T_c)_{exact},(T_c)_{MC}]$ with $(T_c)_{exact}$ and $(T_c)_{MC}$ the theoretical and Monte-Carlo critical temperatures, respectively.  With the success of this benchmarking test, we can move onto the higher $q$ clock models.


\subsection{$q=4$}

The $q=4$ clock model is the dividing line for the appearance of the BKT phase transitions. For the $q>4$ clock models there appears the BKT phase which is characterized by some nonlocal order parameter such as the condensation of vortex/antivortex pair. However, we like to characterize the BKT phase transitions by the patterns of our JSD ``thermometer" for the NN-flowed ensemble of states after $100$ iterations of the initial Monte-Carlo ensemble. 
The result of the NN-flowed ${\rm JSD}[P_{T_{\rm flow}}|| Q_T]$ is shown in Fig. \ref{fig:q4_JSD}. 
The notation for $P_{T_{\rm flow}}$ and $Q_T$, and the three choices of $T_{\rm flow}=$ $T_{\rm flow}$, $T_{\rm mid}$ and $T_{\rm high}$ in subfigures (b) of Fig. \ref{fig:q4_JSD} are the same as for the $q=2$ case. 
In the subfigure (a) of Fig. \ref{fig:q4_JSD},  we plot the color chart for the distributions of ${\rm JSD}[P_{T_{\rm flow}}|| Q_T]$ on the $T$-$T_{\rm flow}$ plane. This plot is for the $q=4$ clock model on a $40\times 40$ lattice. 
We see that the color pattern of Fig.\ref{fig:q4_JSD}(a) is almost uniform along the y-axis, but varies along the x-axis. This implies that all the NN-flowed ensembles of different $T_{\rm flow}$ yield the same JSD profile, that is, there exists a universal fixed-point ensemble of the NN-flowed states. 

More interestingly, the color chart of Fig.\ref{fig:q4_JSD}(a) shows two discontinuities along the x-axis. 
One is around $T=(T_c)_{exact} \approx 1.13$, which is the theoretical critical temperature of the $q=4$ clock mode, and slightly deviates from the Monte-Carlo critical temperature $(T_c)_{MC}$ as indicated by the peak of magnetic susceptibility (yellow line with decorated crosses). 
The second discontinuity is about $T = 1.35$, beyond which the magnetic susceptibility starts to level off. 
In the subfigures (b) of Fig. \ref{fig:q4_JSD} we show the NN-flowed ${\rm JSD}[P_{T_{\rm flow}}|| Q_T]$ for $T_{\rm flow}=T_{\rm low}$ (red), $T_{\rm mid}$ (blue), and $T_{\rm high}$ (green) on a $20\times 20$ (dashes lone) and on a $40 \times 40$ (solid line) lattice. 
Then, we can examine the finite-size effect by comparing these two subfigures. Again, we see that the NN flow drives the JSD curves of Monte-Carlo ensembles of different temperatures to a universal JSD curve with its minimum sitting nearby the narrow window of $[(T_c)_{exact},(T_c)_{MC}]$. 
This is consistent with the result of Fig. \ref{fig:q4_JSD}(a).  
We see that the minimum of the universal JSD on the $40\times 40$ lattice is closer to $(T_c)_{MC}$ than the one of the $20\times 20$ lattice. This indicates the finite-size effect. Thus, after taking care of the finite-size effect, the universal NN-flowed JSD can be used to identify the critical temperature of the $q=4$ clock model as in the $q=2$ case. Lastly, the NN-flowed JSD of the $40\times 40$ lattice also shows a sharp drop in a very small window between $[(T_c)_{exact},(T_c)_{MC}]$.  This can also be thought of as a peculiar feature to indicate the phase transition in addition to using its minimum.

\subsection{$q=5,6,7$}  

Now we will consider the clock models possibly exhibiting the BKT phase transitions, that is, $q>4$. In this subsection we will consider $q=5,6,7$ cases, and in the next subsection the $q=8$ case. From both the analytical and numerical studies \cite{Jose:1977gm,ORTIZ2012780,Tomita2002ProbabilityChangingCA,Krvcmar2016PhaseTO,Chatelain2014DMRGSO,Vanderstraeten:2019frg,Li2020CriticalPO}, it is argued that there is an extended BKT phase in a finite temperature interval $[T_1,T_2]$ for $q>4$ cases. 
However, in \cite{Lapilli_2006} it showed that there exists a so-called {\it extended university} for $q>4$ cases as $T>T_{eu}$ with $T_{eu}$ the onset of thermodynamics collapse and the emerging continuous symmetry. 
Due to the existence of $T_{eu}$, the phase transition at $T_2$ for $q=5,6$ cases turns into the non-BKT one, i.e., no longer a continuous phase transition. This is the reason that we leave the study of the $q=8$ case in the next subsection.

Similar to the $q=2,4$ cases, the resultant NN-flowed ${\rm JSD}[P_{T_{\rm flow}}|| Q_T]$ for $q=5, 6, 7$ on a $40\times 40$ lattice for $T_{\rm flow}=T_{\rm low}$ (red), $T_{\rm mid}$ (blue) and $T_{\rm high}$ (green) of Fig. \ref{fig:JSD_demo} is shown in Fig. \ref{fig:q5_q6_q7flow}. 
We again see that the NN-flowed JSDs approach a universal pattern, which indicates a fixed-point ensemble of states. In the subfigures (a), (b) of Fig. \ref{fig:q5_q6_q7flow} for the $q=5,6$ cases, respectively, we see that the JSD minima sit right on $T=(T_2)_{MC}$, which is the $T_2$ indicated by Monte-Carlo simulation. However, for the subfigure (c) of Fig. \ref{fig:q5_q6_q7flow} for $q=7$ case, the minimum sits at the temperature higher than $(T_2)_{MC}$. Although there is no direct evidence, this could be implied by the non-BKT nature of $(T_2)_{MC}$ due to the existence of $T_{eu} > (T_2)_{MC}$ for $q=5,6$ models. 
For the $q=7$ case, $T_{eu}$ is approximately equal to $(T_2)_{MC}$ so that the phase transition at $T=(T_2)_{MC}$ is BKT type, and the NN flow cannot capture precisely the BKT critical temperature. We will return to this issue for the $q=8$ case in the next subsection. 
 
Finally, in all three cases, as shown in Fig. \ref{fig:q5_q6_q7flow},  there is a peculiar plateau roughly overlapping with the interval $[(T_1)_{MC}, (T_2)_{MC}]$ of the BKT phase, where $(T_{1,2})_{MC}$ are indicated by the vertical light-yellow lines in each subfigure.  This indicates the capability of the NN flow to identify the BKT phase.

\begin{figure}[t!] 
\centering
    \subfigure[ ]{
        \includegraphics[width=0.47\textwidth]{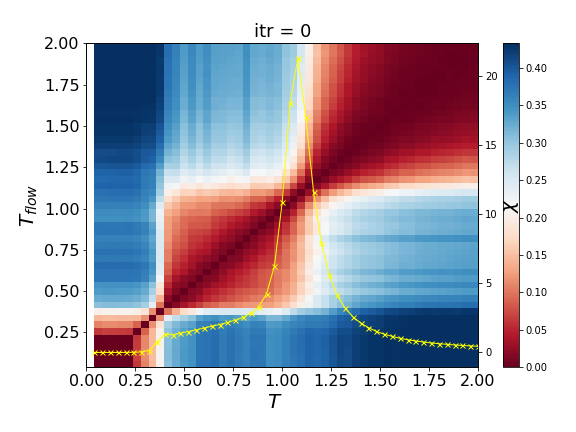} }
    \subfigure[ ]{
        \includegraphics[width=0.47\textwidth]{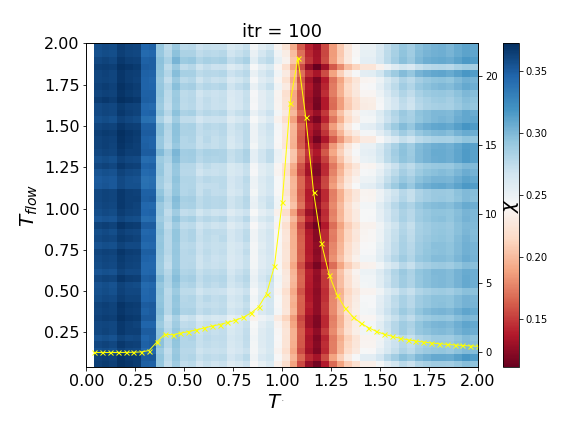} }
\caption{ \small
(a) Color chart of ${\rm JSD}[P_{T_{\rm flow}}||Q_T]$ on the $T$-$T_{\rm flow}$ plane with the PDFs of $P_{T_{\rm flow}}$ and $Q_T$ the ones of mean magnetization for the Monte-Carlo ensembles of the $q=8$ clock model on a $40\times 40$ lattice at temperatures $T_{\rm flow}$ and $T$, respectively. The three (red) blocks coincide with the ordered, BKT, and disordered phases, which are indicated by the superimposed phase diagram of magnetic susceptibility shown by the light-yellow line decorated with crosses. 
(b)  The JSD color chart with the PDF $P_{T_{\rm flow}}$ in 
(a) replaced by the NN-flowed one after 100 iterations. The uniform color distribution along the y-axis indicates a universal fixed-point ensemble, which can capture the Monte-Carlo critical temperatures $(T_{1,2})_{MC}\approx 0.39, 0.93$ by its discontinuity along the x-axis.  }
\label{fig:q8_color}
\end{figure}

\begin{figure}[t!] 
\centering
    \subfigure[ L=20]{
        \includegraphics[width=0.48\textwidth]{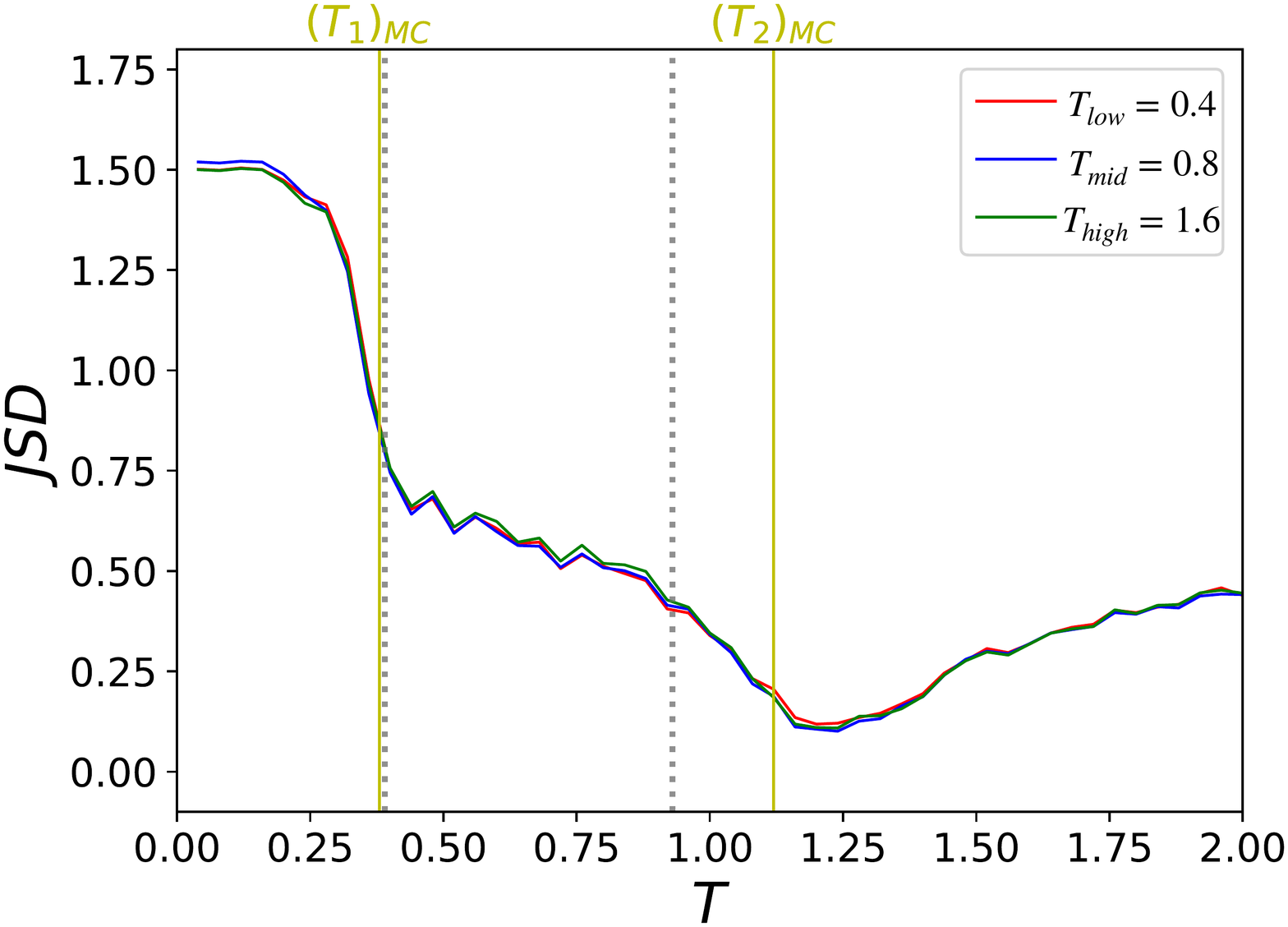} }
    \subfigure[L=40 ]{
        \includegraphics[width=0.48\textwidth]{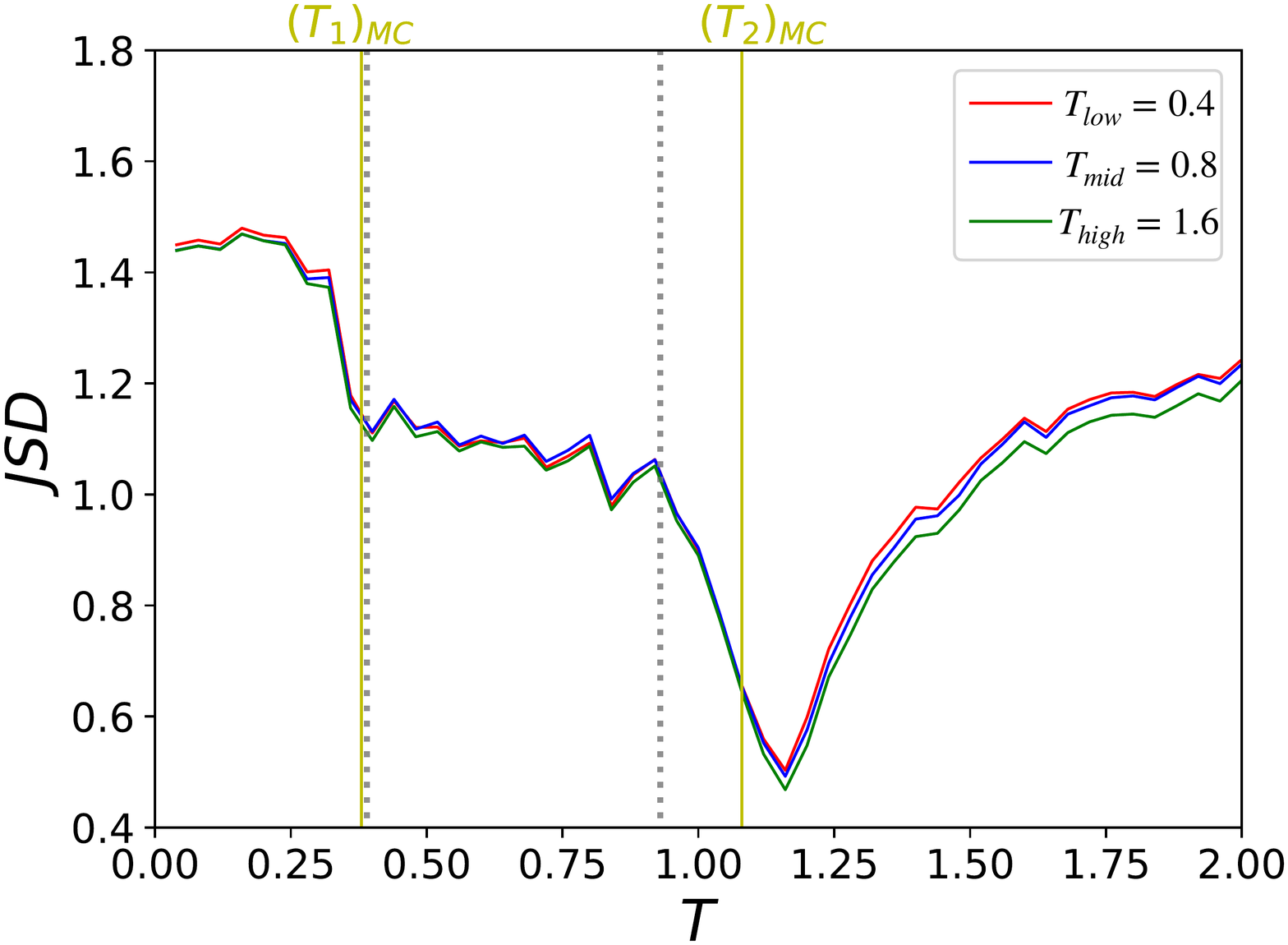} }
\caption{\small NN-flowed ${\rm JSD}[P_{T_{\rm flow}}|| Q_T]$ after 100 iterations for $T_{\rm flow}=T_{\rm low}$ (red), $T_{\rm mid}$ (blue) and $T_{\rm high}$ (green) of Fig. \ref{fig:JSD_demo} for the clock models of $q=8$ (a) on a $20\times 20$ lattice, and (b) on a $40\times 40$ lattice. The results show that NN flow can yield a universal JSD to identify the BKT phase and the associated critical temperatures by taking account of the finite size effect.} 
\label{fig:q8_JSD}
\end{figure}

\subsection{$q=8$}

Now we consider the NN flow of the $q=8$ clock model on $20 \times 20$ and $40 \times 40$ lattices. This model has an extended BKT phase ranging from $(T_1)_{MC}$ to $(T_2)_{MC}$, for which the critical temperatures $(T_{1,2})_{MC}$ of the BKT phase transition are read from the typical phase diagrams of our Monte-Carlo simulation, as shown in Fig. \ref{fig:L40_chi}. Moreover, even there exists the collapse of thermodynamic observables, the phase transition at $(T_2)_{MC}$ is the BKT-type for the $q=8$ clock model because $T_{eu}<(T_2)_{MC}$ \cite{Lapilli_2006}.

We first plot the color charts of ${\rm JSD}[P_{T_{\rm flow}}|| Q_T]$ on the $T$-$T_{\rm flow}$ plane. In Fig. \ref{fig:q8_color}(a), the PDFs $P_{T_{\rm flow}}$ and $Q_T$ are the ones of mean magnetization for the Monte-Carlo ensembles of states at temperatures $T_{\rm flow}$ and $T$, respectively. 
This can be seen as the autocorrelation between finite-temperature Monte-Carlo ensembles.  
Interestingly, three red blocks appear, which correspond to the ordered phase, the BKT phase, and the disordered phase of the $q=8$ clock model. For comparison, in Fig. \ref{fig:q8_color}(b) we replace $P_{T_{\rm flow}}$ in Fig. \ref{fig:q8_color}(a) by the PDF associated with the NN-flowed ensembles after 100 iterations. 
As expected, the color chart is now uniform along the y-axis, indicating a universal fixed-point ensemble of states after NN flow. More interestingly, we see a minor vertical jump near $(T_1)_{MC}\approx 0.39$, which is the lower critical temperature of BKT phase transition. There is no such jump in Fig. \ref{fig:q4_JSD}(a) for the $q=4$ case. On the other hand, there emerges a (red) stripe starting near the second BKT critical temperature $(T_2)_{MC}\approx 0.93$ as indicated by the peak of magnetic susceptibility, but ending when the magnetic susceptibility starts to level off. This kind of (red) stripe also appears in Fig. \ref{fig:q4_JSD}(a) for the $q=4$ case.

Similarly to the cases of $q=5,6,7$, we again show the NN-flowed ${\rm JSD}[P_{T_{\rm flow}}|| Q_T]$ of the $q=8$ clock model for $T_{\rm flow}=T_{\rm low}$ (red), $T_{\rm mid}$ (blue) and $T_{\rm high}$ (green) on a $20\times 20$ lattice in Fig. \ref{fig:q8_JSD}(a), and on a $40\times 40$ lattice in Fig. \ref{fig:q8_JSD}(b). 
The NN-flowed JSDs again approach a unique pattern, indicating a universal fixed-point ensemble of states. 
Unlike the cases of $q=4,5,6$ but similar to the case of $q=7$, the minimum of universal JSD no longer sits on the critical Monte-Carlo temperature $(T_2)_{MC}$ but on a higher one. 
Despite that, the increase of the lattice size does help to sharpen the minimum of the JSD and slightly move it toward $(T_2)_{MC}$. Note that the same effect also moves $(T_2)_{MC}$ to the lower value. By the trend, we expect that with a large enough lattice size, the minimum of the universal JSD could sit closer to $(T_2)_{MC}$ and become sharper. We believe this should be the case for all BKT phase transitions from the BKT phase to the disordered phase, that is, it should hold for $q\ge 7$.

Regarding the BKT phase transition near $(T_1)_{MC}$, the JSDs do show a sharp drop, as in the cases of $q=5, 6, 7$. Moreover, the increase of the lattice size also helps to pin down $(T_1)_{MC}$ more precisely by the endpoint of the sharp drop. We can then conclude that the NN flow can yield a universal JSD pattern that can be used to identify the BKT phase and the associated critical temperatures once the finite size effect is taken into account. 
Therefore, we can infer the extended regime of BKT between $(T_1)_{MC}$ and $(T_2)_{MC}$ from the peculiar pattern of JSD.

\subsection{Properties of the NN-flowed states}

\begin{figure}[t!]
\includegraphics[angle=0,width=0.42
\textwidth]{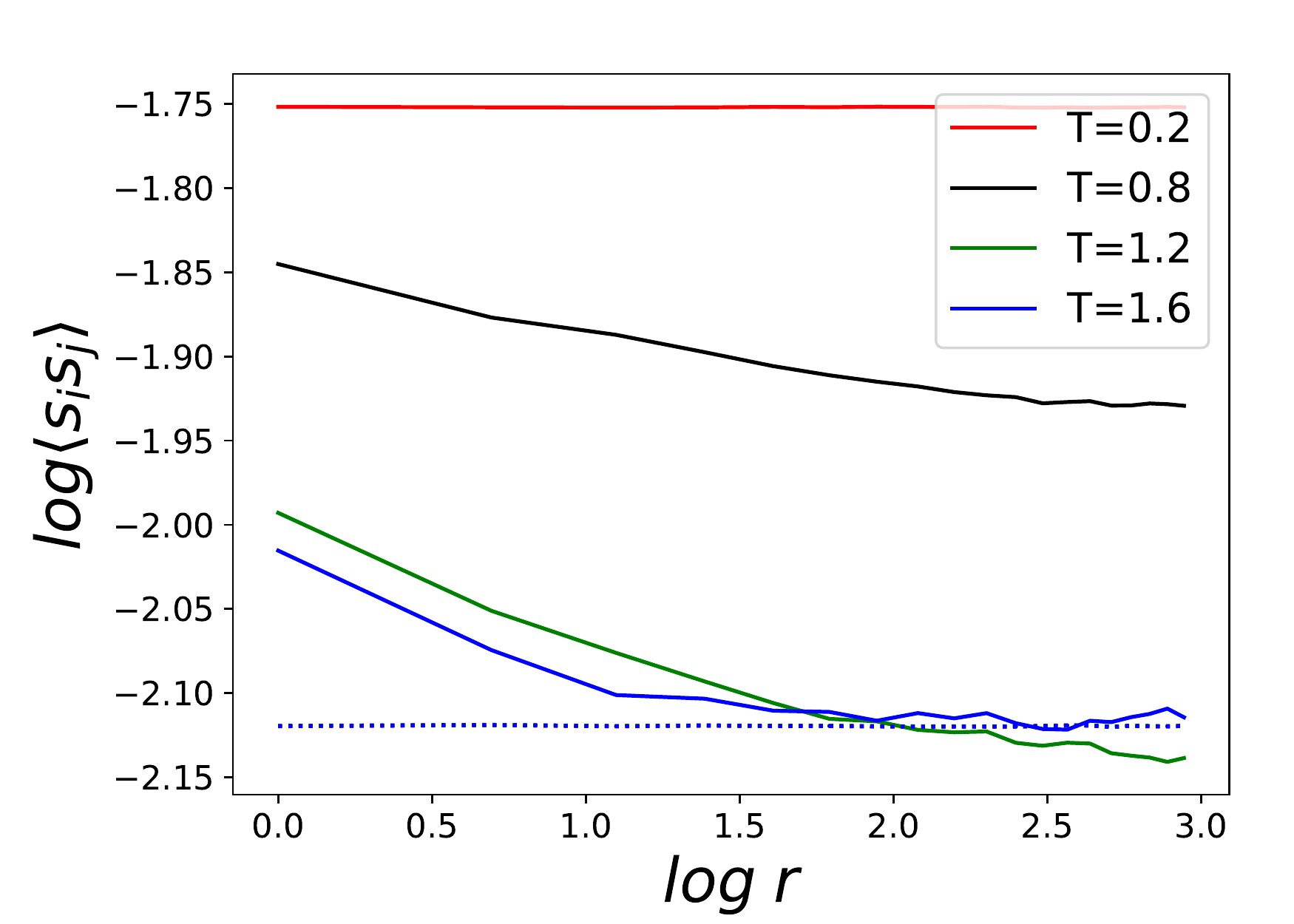}
\caption{\small The correlation of site-spins, $\langle  s_i s_j \rangle$  of the $q=8$ clock model on a $40 \times 40$ lattice as a function of $r=|i-j|$ for the Monte-Carlo ensemble at various temperatures (colored lines) and for the universal NN-flowed ensemble after 100 iterations (dashed lines). We have shown the log-log plots to emphasize the possible power-law behavior, i.e., an approximate straight line of nonzero slope. The $T=0.8$ one (black line) represents the BKT phase and does show an approximate power-law behavior. However, the NN-flowed one is a flat straight line, more like a low-temperature ordered phase, e.g., the red-line one. }
\label{fig:q8_sisj} 
\end{figure}

\begin{figure}[t!]
\includegraphics[angle=0,width=0.42
\textwidth]{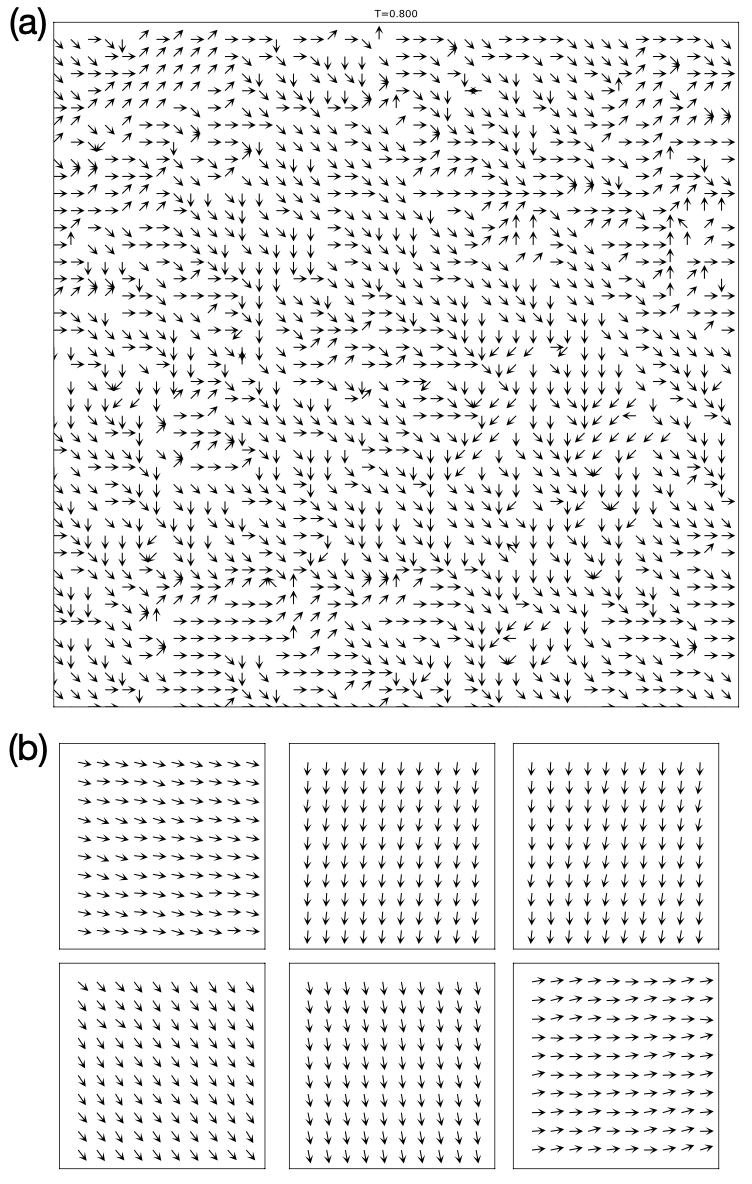}
\caption{\small  (a) The picture of the orientations of the site-spins for a chosen BKT Monte-Carlo state of $q=8$ clock model on a $40\times 40$ lattice with temperature $T=0.8$. This picture shows the typical vortex-antivortex condensation of the BKT phase.  (b) The picture of the orientations of the site-spins for some six chosen NN-flowed states of the same model after $20$ iterations. For simplicity, only part of the lattice is shown, as almost all the site-spins in each NN-flowed state point to the same direction. } 
\label{fig:q8_spin_configuration}
\end{figure}

Our study demonstrates the ability of NN flow to identify BKT phase transitions. This then raises the question if the NN-flowed ensemble fixed-point states can be the physical critical states for the BKT phase transition. Or, does the ensemble of NN-flowed fixed-point states bear any physical implication? In the following, we examine this issue.

To check this issue, we should first recall that our NN-flowed states are no longer the $q$-spin states, but with the site-spin values in the range of $[0,1]$. One can either use this continuous spin values to evaluate some physical quantities or try to discretize it to mimic the $q$-state spin. For the moment, we will just use the continuous site-spin values. 
The first thing to check is to evaluate the spatial correlation function of the site-spins and see if it exhibits power-law-like behaviors, which is the signature of critical states. 
The results for some chosen states of the $q=8$ clock model on a $40\times 40$ lattice are shown in Fig. \ref{fig:q8_sisj}. 
This is a log-log plot so that the power-law behavior should be represented by an approximate straight line of nonzero slope, such as the BKT one at $T=0.8$ shown by the black line. 
However, we see that the NN-flowed one (the dashed line) is just a flat straight line, which is more like the one of the low-temperature ordered phase such as the red-line one. This indicates that the NN-flowed states are more like the low-temperature ordered states, instead of the BKT ones. Such a conclusion remains even when we discretize the site-spins of the NN-flowed fixed-point states. This implies that the NN flow states do not show the features of BKT states, though the NN-flowed JSDs can show a universal pattern which can capture the phase diagram of the BKT phase transitions.

We can explore the above conclusion more directly by showing the orientations of all the site-spins, namely the picture of $0\le \theta_i=2\pi s_i/q \le 2\pi$ for all $i$. 
In Fig. \ref{fig:q8_spin_configuration}(a), we show the picture of the orientations of site-spins for a typical BKT state of the $q=8$ clock model at $T=0.8$, for which we have verified its spin-spin correlation obeys an approximate power-law behavior.  We see that the site spins slowly change their orientations either clockwise or anti-clockwise in the local domains, similar to the vortex and anti-vortex structure. The clockwise domains will usually encounter the anti-clockwise domain afer some extent, expressing the pattern of vortex-antivortex condensation as expected for a BKT state. 
On the other hand, in Fig. \ref{fig:q8_spin_configuration}(a) we show the pictures of orientations of site-spins for six NN-flowed states of the $q=8$ clock model on a $40\times 40$ lattice. Each of them has all the site-spins oriented in the same direction, which is consistent with the result of the flat straight line shown in Fig. \ref{fig:q8_sisj}. For simplicity, we only show part of the lattice. However, different NN-flowed states pick up different orientations for the whole lattice. This seems to imply that each NN-flowed state is in a particular spontaneously symmetry broken state. Despite that, the statistics of the orientation or the mean magnetization from the ensemble of NN-flowed states can yield a universal and peculiar pattern by which the BKT phase transitions can be identified. Therefore, even the NN flow can yield a fixed-point ensemble of states, the component states bear no feature of BKT states so that we cannot use this universal ensemble to evaluate the critical exponents of the phase transitions. This is the essential difference of the NN flow from the RG flow. That is, the statistics of NN-flowed ensemble states can help to identify the BKT phase, but the states themselves carry no feature of the mean-field states. Therefore, we can only take the analogue picture of Fig. \ref{fig:RG_vs_NN} in the literal sense.


\section{Conclusion}
\label{sec:6}

In this paper, we extend the neural flow (NN flow) based on the unsupervised machine learning of the variational autoencoder to study the BKT phase transitions of the $q$-state clock models. Due to the intricate nature of BKT phase, we employ an information-distance measure, the Jensen-Shannon divergence (JSD), as the ``thermometer" to gauge the NN-flowed states with input Monte-Carlo states. We find that the minimum of the JSD can pinpoint the temperatures of the ensembles far more precisely than the usual machine-learning thermometer. With the help of this JSD thermometer, we find that the NN-flowed states from different finite-temperature Monte-Carlo states form a fixed-point ensemble, which can capture the essence of the phase diagrams, such as the BKT phase and the associated critical temperatures of phase transitions. 

However, the NN-flowed states do not bear the feature of BKT states, but each of them looks more similar to a spontaneously symmetry broken states of the low-temperature phase. This implies that the NN flow is not in analogue to the RG flow. Despite that, the statistics of the mean magnetization for the evaluation of the JSDs can indeed capture the essence of the BKT phase diagrams. 

Our results demonstrate the power of machine learning in the study of the topological phase transitions. The realization of the phase diagrams by the NN flow is quite different from the conventional way via coarse-graining or RG flow. It is interesting to explore the deeper statistical structure of the NN-flowed ensemble of states to uncover the mystery of the powerful NN flow method. After the demonstration of this work and the earlier ones \cite{Iso:2018yqu,giataganas1,Koch:2019fxy,Giataganas:2021jqm}, it is also keen to look for the application of NN flow to other physical arenas with the similar critical phenomena, so that NN flow can help to predict the timings or locations of the phase transitions.

\bigskip

\begin{acknowledgments}
We thank Min-Fong Yang for helpful discussions. KKN acknowledges support by the Taiwan Ministry of Science and Technology under Grant no. 110-2112-M-029 -004 and 110-2112-M-029 -006. CYH is supported by the Taiwan Ministry of Science and Technology through Grant No.~108-2112-M-029-006-MY3 and 111-2112-M-029-008. FLL is supported by the Taiwan Ministry of Science and Technology through Grant No.~109-2112-M- 003-007-MY3. 
\end{acknowledgments}

\appendix

\section{More on the BKT phase diagrams of Monte-Carlo simulated configurations}\label{app_a}

\begin{figure*}[t!] 
\centering
\subfigure[]{
        \includegraphics[width=0.325\textwidth]{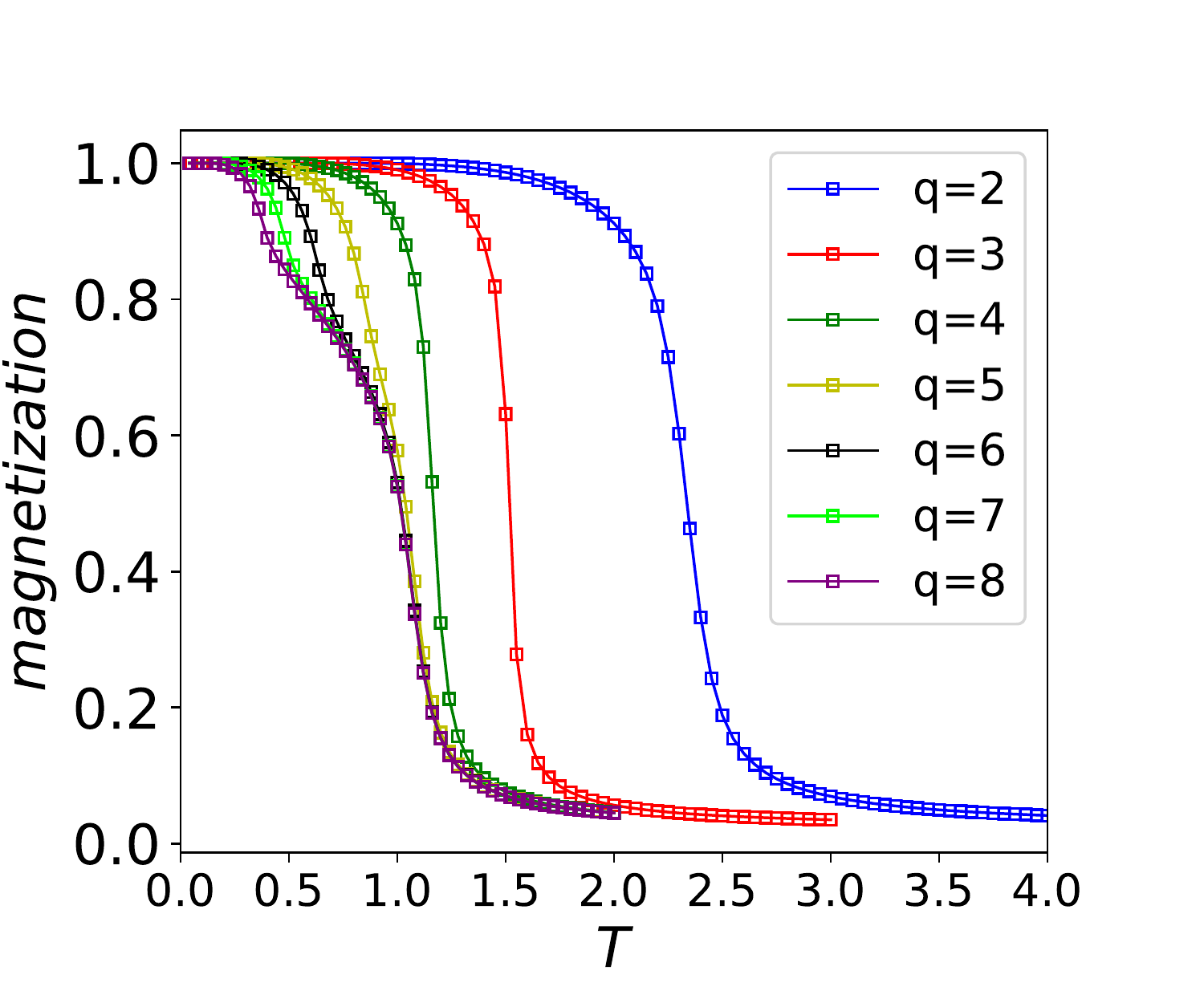}}
\subfigure[]{
        \includegraphics[width=0.325\textwidth]{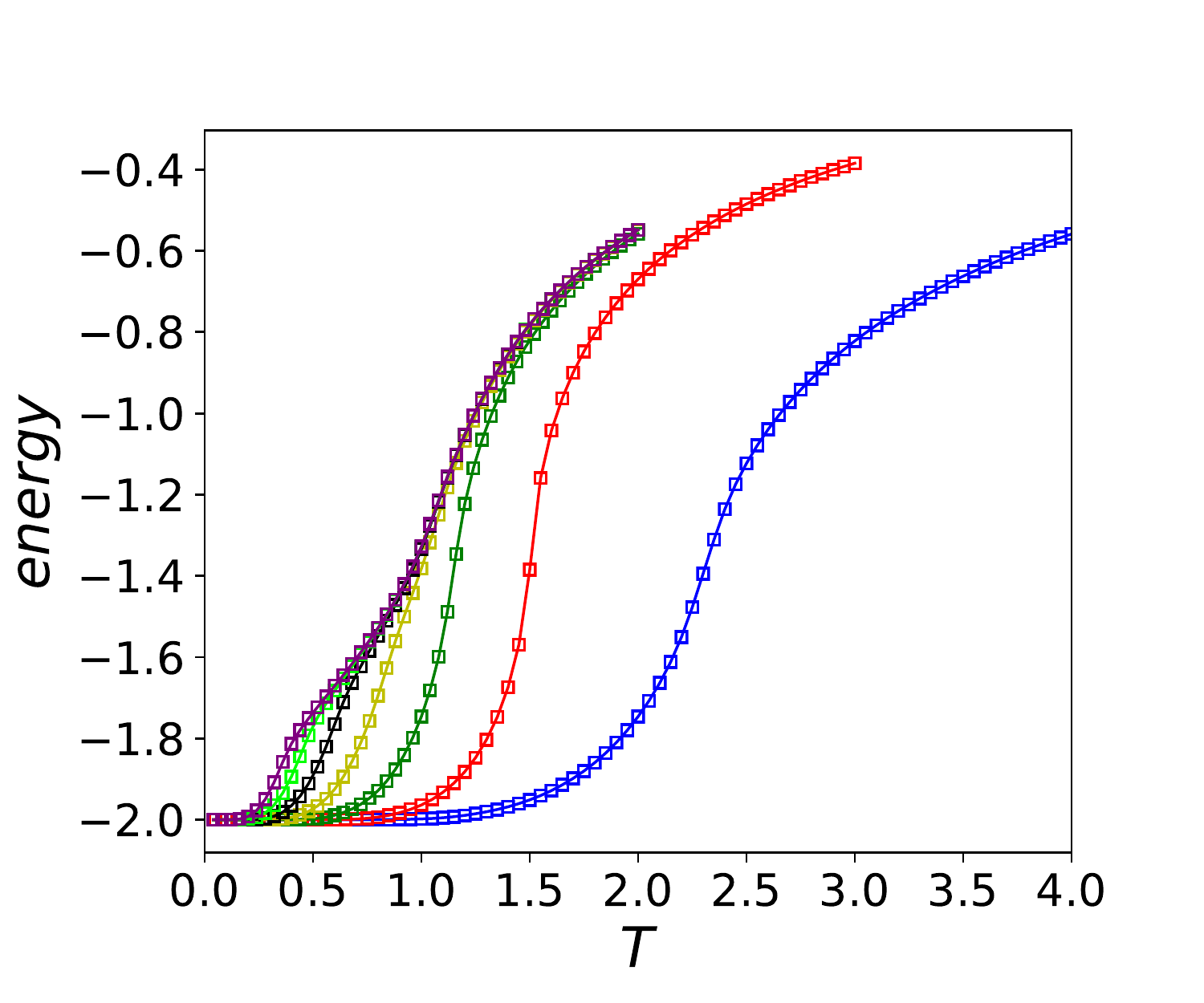}}
\subfigure[]{
        \includegraphics[width=0.325\textwidth]{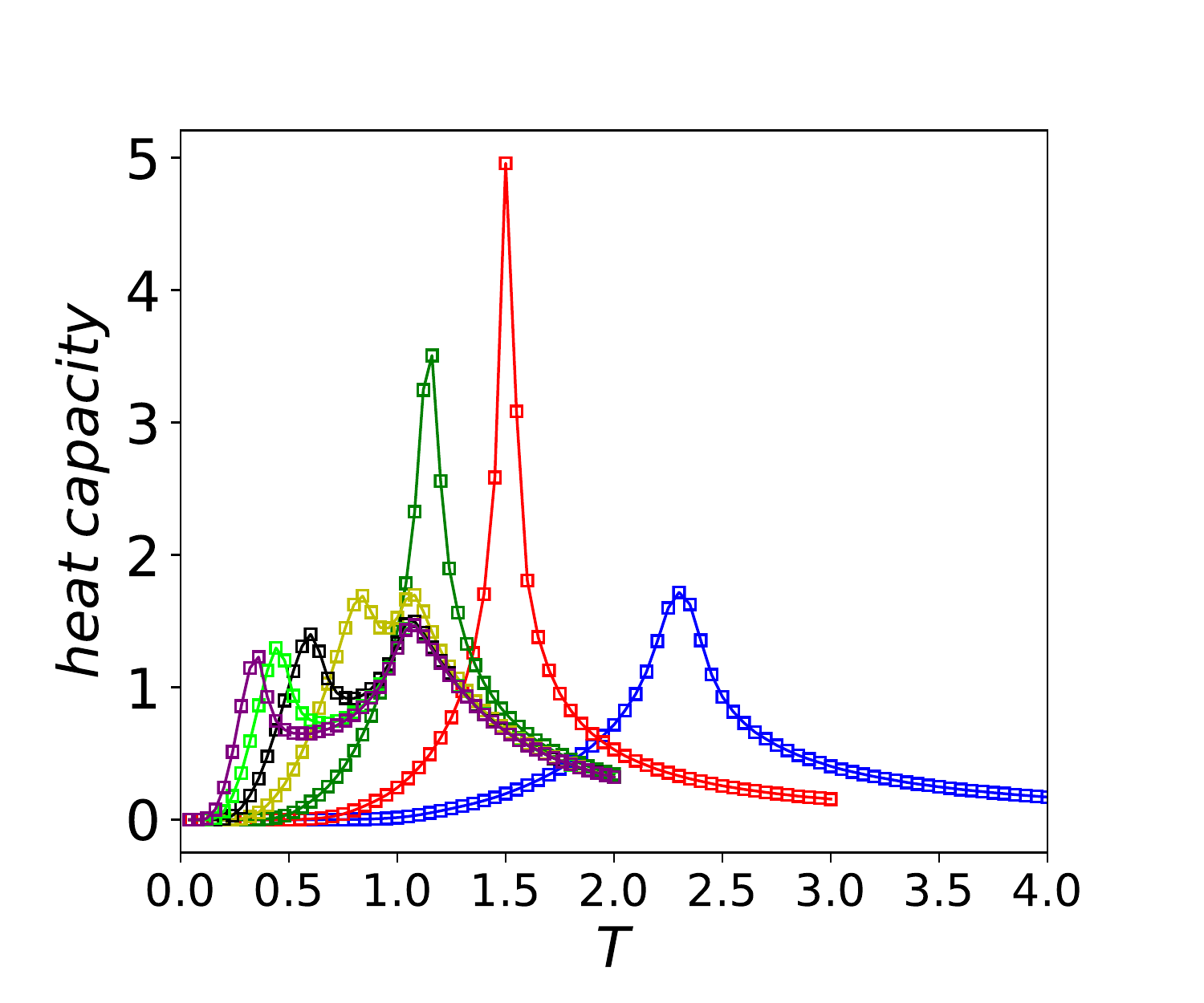}}
\caption{\small  Phase diagrams of $q=2,3,4,5,6,7,8$ clock model  based on the temperature profiles of (a) magnetization, (b) energy, and (c) heat capacity obtained from the Monte-Carlo configurations on a $40 \times 40$ lattice. }
\label{fig:MC_data}
\end{figure*}

For completeness, we show in this appendix the phase diagrams of the $q=2,3,4,5,6,7,8$ clock model  based on the temperature profiles of (a) magnetization, (b) energy, and (c) heat capacity. These results are obtained from the Monte-Carlo configurations of 2000 samples for each temperature bin, and are shown in Fig. \ref{fig:MC_data}. From the results, we find that it is more difficult to identify $T_1$' s of the BKT phase transitions than $T_2$. The latter are more dramatic by a sharp rise/drop (magnetization and energy) or a peak (magnetic susceptibility and energy). the formers, however, show only a small jump/change.

\bibliography{BKT_ref}	

 \end{document}